\documentclass{WileyMSP-template}
\usepackage{multirow}
\usepackage{amsmath,amssymb}
\usepackage{subfig}
\usepackage{caption}
\usepackage{graphicx,subcaption}
\usepackage{comment}
\usepackage{hyperref}
\usepackage{ragged2e}
\usepackage{tikz,xcolor}
\justifying
\begin{document}

\pagestyle{fancy}

\definecolor{lime}{HTML}{A6CE39}
\DeclareRobustCommand{\orcidicon}{
	\begin{tikzpicture}
	\draw[lime, fill=lime] (0,0) 
	circle [radius=0.16] 
	node[white] {{\fontfamily{qag}\selectfont \tiny ID}};
	\draw[white, fill=white] (-0.0625,0.095) 
	circle [radius=0.007];
	\end{tikzpicture}
	\hspace{-2mm}
}

\foreach \x in {A, ..., Z}{\expandafter\xdef\csname orcid\x\endcsname{\noexpand\href{https://orcid.org/\csname orcidauthor\x\endcsname}
			{\noexpand\orcidicon}}
}

\title{Dynamic Continuous Variable Quantum Key Distribution for Securing a Future Global Quantum Network}

\maketitle

\newcommand{\orcidauthorA}{0000-0002-3437-1200} 
\newcommand{\orcidauthorB}{0000-0001-6057-4434} 
\newcommand{\orcidauthorC}{0000-0002-4421-601X} 
\newcommand{\orcidauthorD}{0000-0001-5069-319X} 
\newcommand{\orcidauthorE}{0000-0002-3233-4000} 

\author{Mikhael T. Sayat*,}
\author{Sebastian P. Kish,}
\author{Ping Koy Lam,}
\author{Nicholas J. Rattenbury,}
\author{John E. Cater}



\begin{affiliations}
M. T. Sayat \orcidA{} *\\
Quantum Innovation Centre (Q.InC), Agency for Science Technology and Research (A*STAR), 2 Fusionopolis Way, Innovis \#08-03, Singapore 138634, Republic of Singapore. \\
Department of Physics, Faculty of Science, University of Auckland, Auckland 1010, New Zealand. \\
$\mathrm{mikhael\_sayat@imre.a{-}star.edu.sg}$

Dr. S. P. Kish \orcidB{}\\
Data61, Commonwealth Scientific and Industrial Research Organisation, Sydney, NSW 2015, Australia.

Prof. P. K. Lam \orcidC{}\\
Quantum Innovation Centre (Q.InC), Agency for Science Technology and Research (A*STAR), 2 Fusionopolis Way, Innovis \#08-03, Singapore 138634, Republic of Singapore.\\
ARC Centre for Quantum Computation and Communication Technology, Research School of Physics, Australian National University, Canberra, ACT 2601, Australia.

Dr. N. J. Rattenbury \orcidD{}\\
Department of Physics, Faculty of Science, University of Auckland, Auckland 1010, New Zealand.

Prof. J. E. Cater \orcidE{}\\
Department of Mechanical Engineering, University of Canterbury, Christchurch 8041, New Zealand.

\end{affiliations}


\keywords{Quantum Networks, Quantum Key Distribution, Network Routing, Quantum Communications}

\begin{abstract}
Continuous variable quantum key distribution (CVQKD) is a developing method to secure information exchange in future quantum networks. With the recent developments in quantum technology and greater access to space, a global quantum network secured by CVQKD could be within reach. In this work, the structures of existing QKD networks are analysed, and how they can be fit into a general overarching three-layer QKD network architecture for the endeavour of a global QKD network. Such a network could comprise of different links in fibre and free-space. The finite size limit secret key rates (SKRs) with multidimensional reconciliation were calculated for the different links for which CVQKD can be used in such a network. The results show that CVQKD generally achieves longer distances and larger SKRs in inter-satellite, satellite-to-ground, fibre, and underwater links in descending order. The different links and nodes were classified and secret key distribution was studied as a graph problem. The link capacity, a routing metric for secret key distribution, which considers a dynamic SKR based on dynamic links is presented. Its use in simulated CVQKD networks is presented for the aim of spatiotemporal secret key distribution through a dynamic CVQKD network.

\end{abstract}


\section{Introduction}
Recent developments in quantum technology and access to space positions the development of a global quantum network within reach \cite{Simon2017Towards}. A nascent yet promising method of securing such a network is through continuous variable quantum key distribution (CVQKD). CVQKD uses the continuous amplitude and phase quadrature of light to encode a secret key between communicating parties which is then used to encrypt information \cite{Laudenbach2018Continuous}. Compared to the more established discrete variable QKD, CVQKD offers higher key rates, cheaper detection methods, and is more integrable with existing optical telecommunication networks \cite{Laudenbach2018Continuous}. In addition, CVQKD outperforms discrete variable quantum key distribution (DVQKD) in communicating channels with lower phase noise, and it is more robust against thermal noise \cite{Kish2024Comparison}. The governing performance metric used is the secret key rate (SKR) which represents the size of secret key that can be shared between communicating parties over time.

Many existing national QKD networks exist with different architectures, sizes, protocols, and achievable SKRs \cite{Cao2022Evolution}. The first QKD network was the DARPA quantum network which used the BB84 DVQKD protocol and achieved a SKR of 1 kbit/s \cite{Elliott2002Building,Elliott2005Current}. The Tokyo quantum network consists of six nodes using the decoy state BB84, DPS-QKD, BBM92, and SARG04 DVQKD protocols with a maximum SKR of 304 kbit/s \cite{Sasaki2011Field,Sasaki2011Tokyo}. The SECOQC quantum network exhibits both CVQKD and DVQKD protocols utilising both fibre and free-space channels and achieved a maximum SKR of 17 kbit/s \cite{Dianati2007Architecture,Peev2009Secoqc}. The Madrid quantum network was the first network to be integrated with an existing telecommunication network and used a software-defined network (SDN) architecture which replaced the traditional point-to-point architecture of other networks allowing for easier integration \cite{Martin2019Quantum,Garcia2021Madrid,Martin2024MadQCI}. The SDN architecture controls the whole network as a whole as opposed to point-to-point which controls individual links \cite{wang2019quantum}. The Chinese quantum network is the largest existing network which consists of 166 nodes in fibre and free-space using both CVQKD and DVQKD protocols with a maximum achievable SKR of 235 kbit/s \cite{Chen2021Integrated}. It includes Micius, the first quantum satellite, which achieved BB84 encryption between China and Vienna  with a sifted key rate of 3 to 9 kbit/s over a satellite-to-ground link distance of 1000 to 600 km, respectively \cite{liao2018satellite}.

\begin{figure}[htp!]
    \centering
    \includegraphics[width=0.9\textwidth]{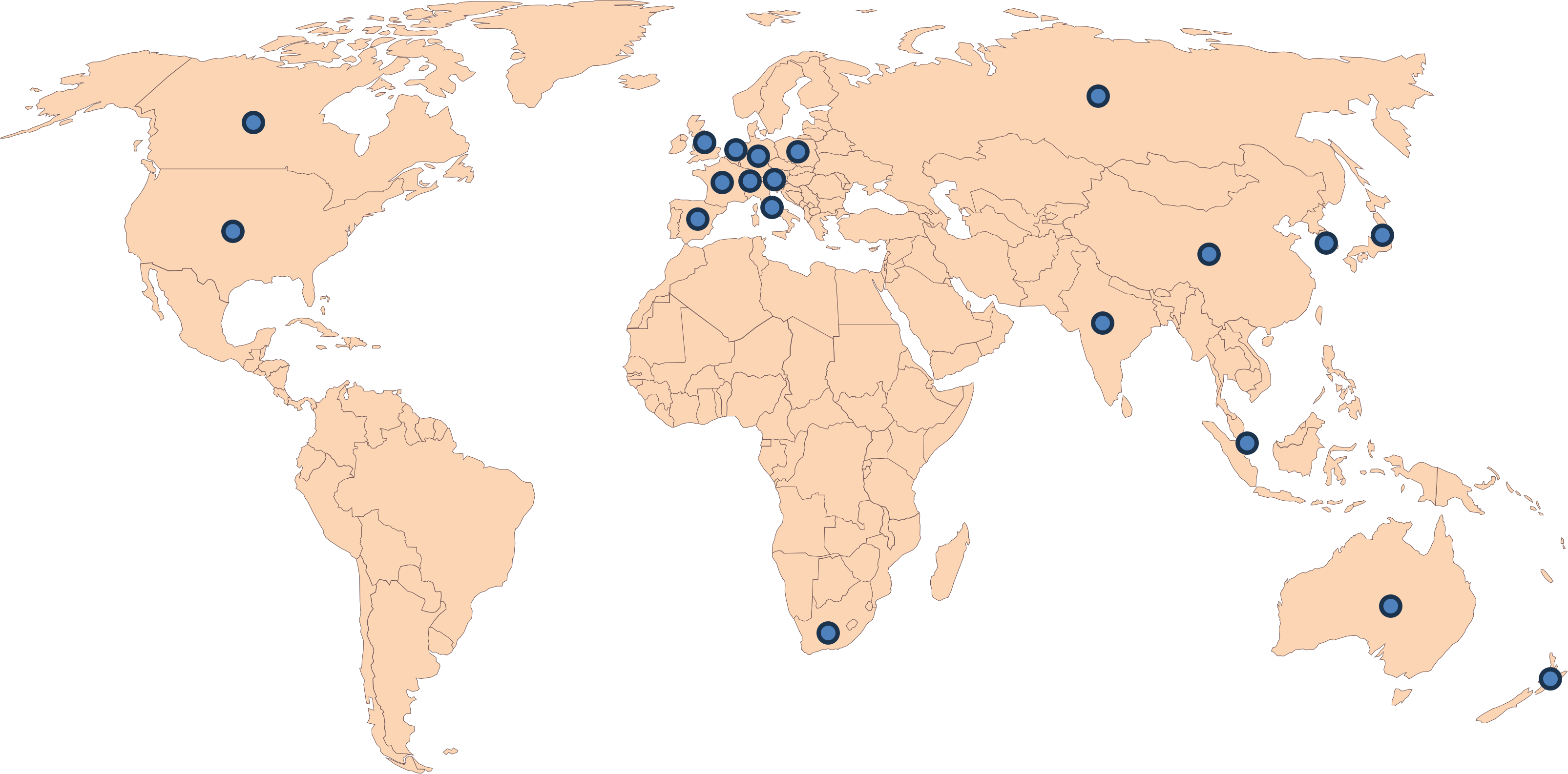}
    \caption[Locations of existing QKD networks, testbeds, demonstrations, and developments]{Locations of existing QKD networks, testbeds, demonstrations, and developments \cite{Cao2022Evolution, mishra2022bbm92, sidhu2021advances,rattenbury2024update}.}
    \label{fig:QKDNetworkLocations}
\end{figure}

Existing networks are predominantly based in fibre and use DVQKD protocols (a summary of significant QKD networks can be found in Table \ref{tab:NetworkSummary} and locations of network developments is shown in Figure \ref{fig:QKDNetworkLocations}). However, it is beneficial to include CVQKD for its advantages and free-space links for spatial diversity. In establishing a secured global quantum network, it is beneficial to be able to use available nodes and links from existing networks for both fibre and free-space QKD. This provides interconnectedness and spatiotemporal diversity whereby the transfer of secret keys can be performed through favourable channels in a dynamic network.

In this paper, we discuss how existing networks can be integrated into an overarching global quantum network architecture. In addition, we collectively present existing models for the feasibility of CVQKD through fibre and available free-space channels which are then used to simulate CVQKD through a dynamic network. Here, we present the link capacity, which is dependent on a dynamic SKR and can be used as a routing metric to choose the optimum pathway in such a network.

\begin{table}[htp!]
\centering
\caption{Features of Existing QKD Networks}
\hspace{-15mm}
\scalebox{1}{
\begin{tabular}{|l|r|r|r|r|r|r|}
\hline
\multicolumn{1}{|c|}{\begin{tabular}[c]{@{}c@{}}QKD \\ Network\end{tabular}} & \multicolumn{1}{c|}{Channels}                                & \multicolumn{1}{c|}{Implementation}                                                             & \multicolumn{1}{c|}{\#Nodes} & \multicolumn{1}{c|}{Max SKR}                           & \multicolumn{1}{c|}{Protocol}                           & \multicolumn{1}{c|}{\begin{tabular}[c]{@{}c@{}}Special\\ Features\end{tabular}}                                      \\ \hline
\begin{tabular}[c]{@{}l@{}}DARPA \\ Quantum \\ Network \cite{Elliott2002Building,Elliott2005Current}\end{tabular}          & \begin{tabular}[c]{@{}r@{}}Fibre,\\ Free-Space\end{tabular}  & \begin{tabular}[c]{@{}r@{}}Optical Switching, \\ Trusted Relay\end{tabular}                     & 10                           & \begin{tabular}[c]{@{}r@{}}10 \\ kbit/s\end{tabular}   & DVQKD                                                   & \begin{tabular}[c]{@{}r@{}}First QKD \\ Network\end{tabular}                                                         \\ \hline
\begin{tabular}[c]{@{}l@{}}Tokyo \\ QKD \\ Network \cite{Sasaki2011Field}\end{tabular}              & Fibre                                                        & Trusted Relay                                                                                   & 6                            & \begin{tabular}[c]{@{}r@{}}304 \\ kbit/s\end{tabular}  & DVQKD                                                   &                                                                                                                      \\ \hline
\begin{tabular}[c]{@{}l@{}}SECOQC\\ Quantum \\ Network \cite{Dianati2007Architecture, Peev2009Secoqc}\end{tabular}          & \begin{tabular}[c]{@{}r@{}}Fibre, \\ Free-Space\end{tabular} & Trusted Relay                                                                                   & 6                            & \begin{tabular}[c]{@{}r@{}}17 \\ kbit/s\end{tabular}   & \begin{tabular}[c]{@{}r@{}}CVQKD, \\ DVQKD\end{tabular} & \begin{tabular}[c]{@{}r@{}}QKD with \\ Micius Satellite\end{tabular}                                                 \\ \hline
\begin{tabular}[c]{@{}l@{}}Cambridge \\ Quantum \\ Network \cite{Dynes2019Cambridge} \end{tabular}      & Fibre                                                        & Trusted Relay                                                                                   & 3                            & \begin{tabular}[c]{@{}r@{}}2.58 \\ Mbit/s\end{tabular} & DVQKD                                                   &                                                                                                                      \\ \hline
\begin{tabular}[c]{@{}l@{}}Bristol \\ Quantum \\ Network \cite{Tessinari2019Field, Joshi2020Trusted}\end{tabular}        & Fibre                                                        & \begin{tabular}[c]{@{}r@{}}Untrusted Relay, \\ SDN\end{tabular}                                 & 8                            & \begin{tabular}[c]{@{}r@{}}83.9 \\ kbit/s\end{tabular} & DVQKD                                                   & \begin{tabular}[c]{@{}r@{}}SDN \\ Implementation\end{tabular}                                                        \\ \hline
\begin{tabular}[c]{@{}l@{}}Madrid \\ Quantum \\ Network \cite{Martin2019Quantum,Garcia2021Madrid}\end{tabular}         & Fibre                                                        & \begin{tabular}[c]{@{}r@{}}Optical Switching, \\ Trusted Relay, \\ SDN\end{tabular}             & 12                           & \begin{tabular}[c]{@{}r@{}}70 \\ kbit/s\end{tabular}   & \begin{tabular}[c]{@{}r@{}}CVQKD, \\ DVQKD\end{tabular} & \begin{tabular}[c]{@{}r@{}}First SDN\\ QKD Network \\ Integrated \\ with Existing \\ Telecom \\ Network\end{tabular} \\ \hline
\begin{tabular}[c]{@{}l@{}}Chinese \\ Quantum \\ Network \cite{Cao2022Evolution, Chen2021Integrated}\end{tabular}          & \begin{tabular}[c]{@{}r@{}}Fibre, \\ Free-Space\end{tabular} & \begin{tabular}[c]{@{}r@{}}Optical Switching, \\ Trusted Relay, \\ Untrusted Relay\end{tabular} & 166                          & \begin{tabular}[c]{@{}r@{}}235 \\ kbit/s\end{tabular}  & \begin{tabular}[c]{@{}r@{}}CVQKD, \\ DVQKD\end{tabular} & \begin{tabular}[c]{@{}r@{}}Largest QKD \\ Network and \\ is Integrated \\ with Space\end{tabular}                    \\ \hline
\end{tabular}
}
\label{tab:NetworkSummary}
\end{table}

\newpage
\section{Integrating QKD Networks}
\begin{figure}[!htp]
   \centering
   \includegraphics[width=0.7\textwidth]{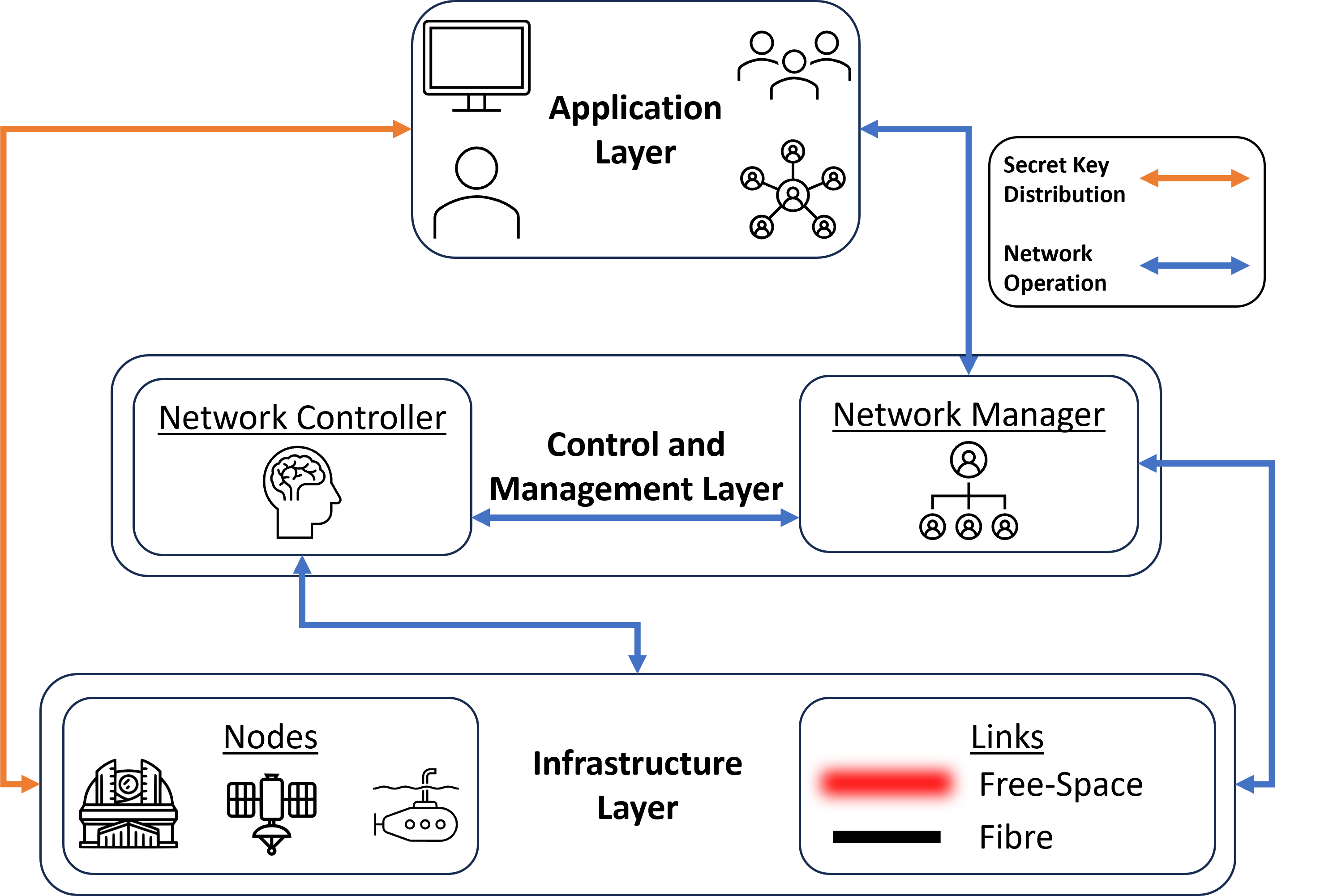}
   \caption[General architecture of a QKD network]{General architecture of a QKD network, consisting of the infrastructure layer (nodes, links, QKD-related devices), control and management layer (network controller and manager), and application layer (users) \cite{sayat2024towards}.}
   \label{fig:NetworkArchitecture}
\end{figure}

Multiple QKD networks can be connected into one overarching network architecture, such as that shown in Figure \ref{fig:NetworkArchitecture} which comprises of three layers \cite{sayat2024towards}. The infrastructure layer consists of all the nodes, links between nodes, and any QKD-related devices required for the distribution and storing of secret keys. The control and management layer consists of the network controller and network manager. The network controller calibrates and controls the activation of nodes, and the link connections between them, for QKD operation. The network manager monitors the QKD nodes and links, keeping track of secret key and channel parameters, as well as supervising the network controller's calibration and activation of the infrastructure layer. 

The application layer represents the end-users of QKD who require a secret key. Users request secret keys from the network manager which then queries a corresponding node for a secret key. Once a secret key is available, the network manager instructs the network controller to activate the corresponding node to send the secret key to the user. It should be noted that the control and management layer exists only to activate, calibrate, and monitor the infrastructure layer. The distribution of secret keys occurs between the nodes and the application layer where the contents of the secret key remain undisclosed to the control and management layer.

In constructing a global quantum network, different national networks may need to adhere to a global QKD network architecture such as that shown in Figure \ref{fig:NetworkArchitecture}. Operations and layer structures would need to be made compatible. The use of an SDN controller would allow easier integration, as demonstrated by the Madrid quantum network \cite{Martin2019Quantum}. 

Using the general three-layer QKD network architecture, the infrastructure layer would consist of the
\begin{itemize}
    \item Nodes: OGS, satellites, submarines, high-altitude platforms, and
    \item Links: fibre, satellite-to-ground, submarine-to-ground, satellite-to-submarine, inter-satellite, inter-submarine.
\end{itemize}
The network manager in the control and management layer would monitor the channel parameters (transmittance, excess noise) based on the environment conditions inherent to each link e.g. visibility, refractive index structure parameter, chlorophyll concentration, link distances etc.

Although a general QKD network could consist of the three aforementioned layers (Figure \ref{fig:NetworkArchitecture}), other networks (proposals and experimental) have adopted four layers \cite{Maeda2009Technologies, Cao2017Key, Cao2017Resource, Tajima2017Quantum, Zhao2018Resource, Cao2019Kaas}, five layers \cite{Chen2021Integrated}, and six layers \cite{ITU-T2019}. The additional layers arise from further division of the three general layers (infrastructure, control and management, and application). If a global quantum network used the general three-layer QKD network architecture, the different number of layers in existing QKD networks \cite{Cao2022Evolution} would need to be integrated into and made compatible with the three layers. Table \ref{tab:integratedNetworks} shows the different layers of existing QKD networks and where they fit into the general three-layer architecture.

\begin{table}[htp!]
\centering
\caption{QKD Networks in the Three Layer General Network Architecture}
\begin{tabular}{|l|rrr|}
\hline
\multirow{2}{*}{}                                                       & \multicolumn{3}{c|}{\textbf{General Network Architecture Layer}}                                                                                                                                                                                                                                                                                                               \\ \cline{2-4} 
                                                                        & \multicolumn{1}{c|}{\begin{tabular}[c]{@{}c@{}} \textbf{Application} \\ \textbf{Layer}\end{tabular}}                          & \multicolumn{1}{c|}{\begin{tabular}[c]{@{}c@{}}\textbf{Control and} \\ \textbf{Management Layer}\end{tabular}}                                                      & \multicolumn{1}{c|}{\begin{tabular}[c]{@{}c@{}}\textbf{Infrastructure} \\ \textbf{Layer}\end{tabular}}                 \\ \hline
\begin{tabular}[c]{@{}l@{}}DARPA \\ Quantum \\ Network \cite{elliot2007final} \end{tabular}     & \multicolumn{1}{r|}{IKE Daemon}                                                                            & \multicolumn{1}{r|}{\begin{tabular}[c]{@{}r@{}}Quantum Protocol \\ Daemon, Optical \\ Process Control\end{tabular}}                               & \begin{tabular}[c]{@{}r@{}}Physical \\ Layer (Tx/Rx)\end{tabular}                                    \\ \hline
\begin{tabular}[c]{@{}l@{}}Tokyo QKD \\ Network \cite{Sasaki2011Field} \end{tabular}            & \multicolumn{1}{r|}{\begin{tabular}[c]{@{}r@{}}Communication \\ Layer\end{tabular}}                        & \multicolumn{1}{r|}{\begin{tabular}[c]{@{}r@{}}Key \\ Management \\ Layer\end{tabular}}                                                           & Quantum Layer                                                                                        \\ \hline
\begin{tabular}[c]{@{}l@{}}SECOQC QKD \\ Network \cite{Peev2009Secoqc} \end{tabular}           & \multicolumn{1}{r|}{\begin{tabular}[c]{@{}r@{}} Data Plane\end{tabular}} & \multicolumn{1}{r|}{\begin{tabular}[c]{@{}r@{}} Secret's Plane\end{tabular}}                          & \begin{tabular}[c]{@{}r@{}}Quantum Plane\end{tabular} \\ \hline
\begin{tabular}[c]{@{}l@{}}Cambridge \\ Quantum \\ Network \cite{Dynes2019Cambridge} \end{tabular} & \multicolumn{1}{r|}{Application Layer}                                                                     & \multicolumn{1}{r|}{\begin{tabular}[c]{@{}r@{}}Network Key \\ Delivery Layer\end{tabular}}                                                        & Quantum Layer                                                                                        \\ \hline
\begin{tabular}[c]{@{}l@{}}Bristol \\ Quantum \\ Network \cite{Tessinari2019Field}\end{tabular}   & \multicolumn{2}{r|}{SDN Plane}                                                                                                                                                                                                                                 & Data Plane                                                                                           \\ \hline
\begin{tabular}[c]{@{}l@{}}Madrid \\ Quantum \\ Network \cite{Garcia2021Madrid}\end{tabular}    & \multicolumn{3}{r|}{SDN}                                                                                                                                                                                                                                                                                                                                              \\ \hline
\begin{tabular}[c]{@{}l@{}}Chinese \\ Quantum \\ Network \cite{Chen2021Integrated}\end{tabular}     & \multicolumn{1}{r|}{Application Layer}                                                                     & \multicolumn{1}{r|}{\begin{tabular}[c]{@{}r@{}}Classical Logical \\ Layer, Classical \\ Physical Layer, \\ Quantum Logical \\ Layer\end{tabular}} & \begin{tabular}[c]{@{}r@{}}Quantum \\ Physical Layer\end{tabular}                                    \\ \hline
\end{tabular}
\label{tab:integratedNetworks}
\end{table}

\section{Secret Key Rate}
\label{sec:SecretKeyRate}
The secret key rate in the asymptotic limit, which assumes that an infinite number of symbols are sent, can be defined as \cite{Laudenbach2018Continuous,Denys2021Explicit}:

\begin{equation}
    \mathrm{SKR}_{\mathrm{asy}} = \beta I_{AB} - S_{BE},
    \label{SKR_asy}
\end{equation}
where $\beta$ is the reconciliation efficiency based on the information reconciliation step during classical post processing. $I_{AB}$ is the mutual information between Alice and Bob, and is the maximum amount of information that can be shared depending on the protocol and channel parameters. $S_{BE}$ is the estimated upper bound on the information about the key known by Eve, and is known as the ``Holevo Bound". Since $S_{BE}$ is an upper bound, Equation \ref{SKR_asy} is the lower bound on the SKR.

All information required for the calculation of the SKR in a CVQKD system is contained in the system's covariance matrix, which is defined as \cite{Denys2021Explicit}:

\begin{equation}
    \Gamma = \begin{bmatrix}
(V_A + 1)\mathbf{I} & Z\mathbf{\sigma_Z}\\
Z\mathbf{\sigma_Z} & (1 + TV_A + T\xi)\mathbf{I} 
\end{bmatrix} = \begin{bmatrix}
V\mathbf{I} & Z\mathbf{\sigma_Z}\\
Z\mathbf{\sigma_Z} & W\mathbf{I}    
\end{bmatrix},
\label{CovarianceMatrix}
\end{equation}
where $T$ is the transmittance and $\xi$ is the excess noise (in SNU). The transmittance is the inverse of the overall attenuation experienced by a laser between Alice and Bob due to loss. The excess noise comes from unaccounted noise detected by Bob. $\mathbf{I}$ is the identity matrix and $\mathbf{\sigma_{z}}$ is the 3rd Pauli matrix, $\begin{bmatrix} 1 & 0 \\ 0 & -1 \end{bmatrix}$. $Z$ is the correlation coefficient which depends on the specific modulation scheme (refer to \cite{Denys2021Explicit} for calculation). The parameters in Equation \ref{SKR_asy} are calculated depending on the coherent state detection measurement \cite{Laudenbach2018Continuous,Denys2021Explicit}. Homodyne detection is when only one quadrature is measured at a time. Heterodyne detection occurs when both quadratures are measured simultaneously.

The mutual information is defined as:


\begin{equation}
    I_{AB_\mathrm{hom}} = \frac{1}{2}\log_2(1 + \mathrm{SNR}),
    \label{IAB_hom}
\end{equation}
for homodyne detection, and

\begin{equation}
    I_{AB_\mathrm{het}} = \log_2(1 + \mathrm{SNR}),
    \label{IAB_het}
\end{equation}
for heterodyne detection, where
\begin{equation}
    \mathrm{SNR} = \frac{TV_A}{2 + T\xi}.
\end{equation}
Here, the mutual information for heterodyne detection is doubled as both quadratures are measured simultaneously.

The Holevo Bound is defined as \cite{Denys2021Explicit}:

\begin{equation}
    S_{BE} = g\left(\frac{\lambda_1 - 1}{2}\right) + g\left(\frac{\lambda_2 - 1}{2}\right) - g\left(\frac{\lambda_3 - 1}{2}\right),
\end{equation}
where $g(x) = (x+1)\log_2(x+1) -x\log_2(x)$ and $\lambda_1$ and $\lambda_2$ are the symplectic eigenvalues of the covariance matrix, $\Gamma$. $\lambda_3$ is defined as \cite{Denys2021Explicit}

\begin{equation}
    \lambda_{3, \mathrm{hom}} = \sqrt{V\left(V - \frac{Z^2}{W}\right)},
\end{equation}
for homodyne detection, and

\begin{equation}
    \lambda_{3, \mathrm{het}} = V - \frac{Z^2}{W + 1}.
\end{equation}

The SKR in the finite size limit, assuming a finite number of symbols are sent is defined as \cite{Johnson2017Problem, Jeong2022Rate}:

\begin{equation}
    \mathrm{SKR_\mathrm{fin}} = f \left[(1 - \mathrm{FER})\beta I_{AB} - S_{BE} -\delta n_{\mathrm{privacy}} \right],
    \label{SKR_fin_2}
\end{equation}
where $f$ is the laser repetition rate (the amount of symbols transmitted per second), $\mathrm{FER}$ is the frame error rate (the rate at which errors arise in one block of the finite key), $v$ is the fraction of symbols sent that are excluded for channel parameter estimation, and $\delta n_{\mathrm{privacy}}$ is the fraction of symbols sent that are excluded for estimating the information gained by Eve that reflects the validity of estimated channel parameters in determining the Holevo Bound. 

$\delta n_{\mathrm{privacy}}$ is calculated as \cite{sayat2024satellite}:

\begin{equation}
     \delta n_{\mathrm{privacy}} = \frac{(d + 1)^2}{\sqrt{N}} +\frac{ 4(d+1)\sqrt{\log_2(\frac{2}{\epsilon_s})}}{\sqrt{N}} + \frac{2\log_2(\frac{2}{\epsilon^2 \epsilon_s})}{\sqrt{N}} + \frac{\frac{4\epsilon_s d}{\epsilon \sqrt{N}}}{\sqrt{N}},
     \label{privacy}
\end{equation}

where $d$ is a discretisation parameter, $\epsilon_s$ is a smoothing parameter, $\epsilon$ is a security parameter representing the probability that the key is not secret, and $N$ is the total number of symbols sent between Alice and Bob. The details of these different parameters can be found in previous work \cite{Kish2020Feasibility, Hosseinidehaj2020Finite, Leverrier2015Composable, Lupo2018Continuous}. 

Table \ref{tab:securityparameters} shows the security parameter values used for finite size limit SKR calculations in \cite{sayat2024satellite} for CVQKD with security against collective attacks \cite{pirandola2021limits,Pirandola2021Satellite}.

\begin{table}[htp!]
    \centering
    \caption{Security Parameter Values for Finite Size Limit SKR Calculations}
    \begin{tabular}{|c|c|}
        \hline
        \textbf{Security Parameter} & Value \\
        \hline
        \hline
        $d$                     & 5   \\
        $\epsilon_s$             & $2\times 10^{-10}$  \\
        $\epsilon$               & $1\times 10^{-9}$  \\
        $N$                     & $1\times 10^{11}$ \\ 
        \hline
    \end{tabular}
    \label{tab:securityparameters}
\end{table}

Previous experimental CVQKD demonstrations have used laser repetition rates between 10 and 500 MHz \cite{Wang202125, Hirano2017Implementation, Huang2013300, Ren2021Demonstration}. Unless otherwise stated, a laser repetition rate of $f = 50 \;\mathrm{MHz}$ has been selected for the calculation of finite size limit SKRs, which is near the lower end of the 10--500 MHz range of laser repetition rates used in successfully demonstrated CVQKD, to show modest and more achievable finite size limit SKRs that do not require specialised lasers.

The reconciliation efficiency, $\beta$, in the finite size limit depends on the information reconciliation method such as multilevel coding and multistage decoding (MLC-MSD) or multidimensional (MD) reconciliation, which can be used alongside low density parity check (LDPC) codes and depends on the SNR \cite{Mani2021Multiedge} which is defined as \cite{sayat2024satellite}:

\begin{equation}
    \mathrm{SNR} = 10\log_{10}\left({\frac{TV_A}{2 + T\xi}}\right) \;\; [\mathrm{dB}].
\end{equation}

The dependency of $\beta$ on the SNR has been empirically determined as \cite{sayat2024satellite, Mani2021Multiedge}:

\begin{equation}
    \beta_{\mathrm{MLC-MSD/MD}} = c_1^{c_2\mathrm{SNR}} - c_3^{c_4\mathrm{SNR}},
    \label{eqn:Beta_SNR}
\end{equation}
where the coefficients, $c_i$, can be found in Table \ref{tab:beta} for MLC-MSD and MD reconciliation.

\begin{table}[htp!]
    \centering
    \caption{Coefficients of $\beta$ in the Finite Size Limit \cite{sayat2024satellite}}
    \begin{tabular}{|c|r|r|}
        \hline
        \textbf{Coefficient} & MLC-MSD & MD \\
        \hline
        \hline
        $c_1$                     & 0.9655  & $8.250\times10^{-2}$ \\
        $c_2$                     & $1.507\times10^{-4}$ & 0.1834 \\
        $c_3$                     & $-4.696\times10^{-2}$ & 0.9821 \\
        $c_4$                     & -0.2238 & $-2.815\times10^{-5}$ \\ 
        \hline
    \end{tabular}
    \label{tab:beta}
\end{table}

In addition, the FER depends on the SNR and has been empirically defined as \cite{sayat2024satellite,Mani2021Multiedge}:

\begin{equation}
    \mathrm{FER} = \frac{1}{2}\left[1 + m_1\arctan(m_2\mathrm{SNR} + m_3)\right],
    \label{eqn:FER_SNR}
\end{equation}
where $m_1 = 0.8218$, $m_2 = -19.46$, and $m_3 = -298.1$. Note that both $\beta$ and the FER asymptotically approach 1 as the SNR decreases \cite{Laudenbach2018Continuous,sayat2024satellite}. In addition, Equations \ref{eqn:Beta_SNR} and \ref{eqn:FER_SNR} are only valid for values between 0 and 1 \cite{sayat2024satellite}.

\section{Nodes and Links for CVQKD}
\label{sec:QKDNetworks}
\begin{figure}[htp!]
    \centering
    \includegraphics[width=0.8\textwidth]{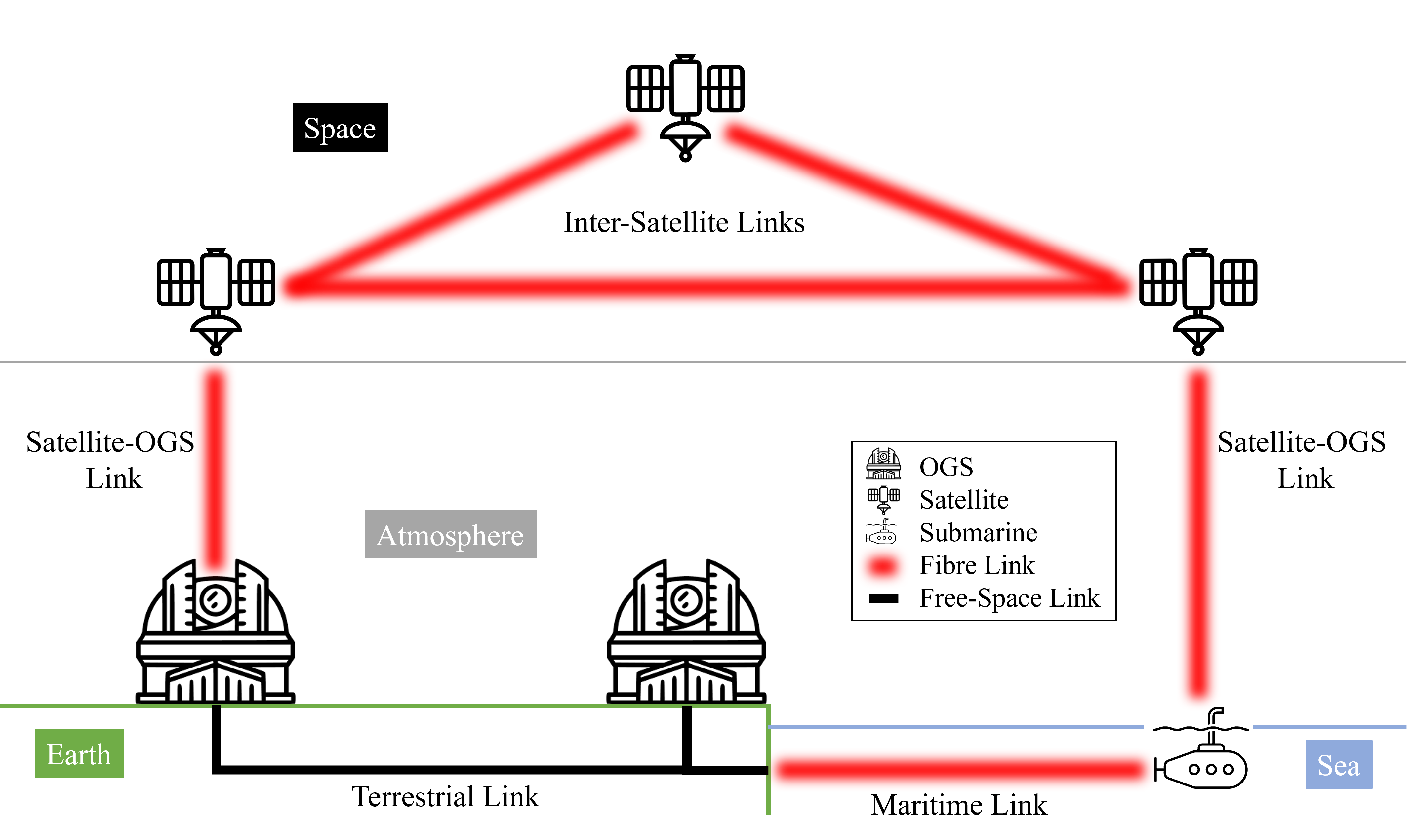}
    \caption[Summary of different nodes and links that could be used in a global quantum network]{Summary of different nodes and links that could be used in a global quantum network for fibre-based and free-space CVQKD \cite{sayat2024towards}.}
    \label{fig:GlobalQuantumCVQKDNetwork}
\end{figure}

Figure \ref{fig:GlobalQuantumCVQKDNetwork} shows nodes and links that could be used in establishing a CVQKD enhanced global quantum network. The links can be characterised by losses and excess noise that would influence the performance of CVQKD protocols. The nodes would comprise of technologies capable of supporting these links.

For links with multiple sources of loss ($A_i$), the overall transmittance is defined as:
\begin{equation}
    T = \prod_{i=1}^n \frac{1}{A_i},
\end{equation}
where $n$ is the number of sources of loss. Note that $A$ is defined as $\frac{P_T}{P_R}$ where $P_T$ is the transmitted power and $P_R$ is the received power. In general, it is found that for a constant value of noise, larger and more positive SKRs for longer link distances can be achieved in inter-satellite links followed by satellite-to-ground, fibre, and underwater links in order; this can be seen in Appendix A where general trends are also presented.

\subsection{Fibre Links}
Terrestrial fibre links connect terrestrial nodes such as optical ground stations and are typically embedded in the ground. The transmittance attributed to using a laser in standard fibre is defined as\cite{Laudenbach2018Continuous, Denys2021Explicit}:
\begin{equation}
    T_{\mathrm{fibre}} = 10^{-0.02d},
    \label{FibreLoss}
\end{equation}
and depends on the link distance, $d$ in km, and the fibre attenuation factor. Here, the attenuation factor is 0.02 dB/km \cite{Denys2021Explicit}. It should be noted that investigation of the use of low loss fibres for CVQKD is an ongoing field of development \cite{Zhang2020Long,hajomer2024ong}. However, for modelling fibre transmissions Equation \ref{FibreLoss} is used. In fact, the relatively large losses from fibre is a significant reason for moving towards free-space optical communications and QKD \cite{Hosseinidehaj2018Satellite}.

\subsection{Underwater}
Underwater maritime links can be defined as a linear attenuation model, assuming direct line of sight, by calculating the radiative transfer equation following Beer's Law \cite{Zhao2022Monte}. The transmittance is defined as \cite{Zhao2022Monte, Gabriel2011Channel, Kraemer2019Monte}:

\begin{equation}
    T_{\mathrm{underwater}} = \exp\left({-c(\lambda)z}\right),
    \label{T_Maritime}
\end{equation}
where $z$ is the link distance and $c(\lambda)$ is a beam extinction coefficient that is dependent on the wavelength as well as optical absorption and scattering underwater. The beam extinction coefficient is defined as \cite{Zhao2022Monte, Gabriel2011Channel, Kraemer2019Monte}:

\begin{equation}
    c(\lambda) = a(\lambda) + b(\lambda),
\end{equation}
where $a(\lambda)$ is the absorption coefficient and $b(\lambda)$ is the scattering coefficient. 

The absorption coefficient is defined as \cite{Xu2016Underwater,Kong2017Security}:

\begin{equation}
    a(\lambda) = \left[ a_w(\lambda) + 0.06a_c(\lambda)C^{0.65}  \right] \left[  1 + 0.2\exp(-0.014(\lambda - 440)) \right],
\end{equation}
where $a_w$ is the absorption coefficient of pure water, $a_c$ is the chlorophyll-specific absorption coefficient, and $C$ is the chlorophyll concentration (data are available in \cite{Morel1977Analysis, Prieur1981Optical}). 

The scattering coefficient is defined as:

\begin{equation}
    b(\lambda) = 0.3\frac{550}{\lambda}C^{0.62}.
    \label{Scat_Maritime}
\end{equation}

Although Equations \ref{T_Maritime}--\ref{Scat_Maritime} define the transmittance for all wavelengths, it should be noted that absorption at near-mid infrared wavelengths (1064--1550 nm) is significant. This can be seen from extrapolation of published absorption coefficient data \cite{Morel1977Analysis, Prieur1981Optical}. Underwater CVQKD can be operated in the near ultraviolet to visible spectra (400-700 nm). Typical values of $a(\lambda)$ and $b(\lambda)$ are shown in Table \ref{tab:MaritimeCoefficients} for different water purities at $\lambda = 520$ nm \cite{Kong2017Security}.

\begin{table}[htp!]
    \centering
    \caption{Absorption and Scattering Coefficients for Water at 520 nm}
    \begin{tabular}{|l|c|c|c|}
        \hline
        \textbf{Water Type} & a ($\mathrm{m^{-1}}$) & b ($\mathrm{m^{-1}}$) & \textbf{c ($\mathbf{\mathrm{m^{-1}}}$)} \\
        \hline
        \hline
        Pure Sea Water & 0.0405 & 0.0025 & 0.043 \\
        Clear Ocean Water & 0.114 & 0.037 & 0.151\\
        Coastal Ocean Water & 0.179 & 0.219 & 0.398 \\
        Turbid Harbour Water & 0.366 & 1.824 & 2.190 \\
        \hline
    \end{tabular}
    \label{tab:MaritimeCoefficients}
\end{table}

\subsection{Inter-Satellite Links}
Inter-satellite free-space links connect satellites in space. The satellites can be in different orbital altitudes from Earth \cite{orbitsESA}:
\begin{itemize}
    \item Low Earth Orbit (LEO): 160-1000~km,
    \item Geosynchronous equatorial orbit (GEO): 35,786 km, and
    \item Medium Earth Orbit (MEO): between LEO and GEO.
\end{itemize}

In space, absorption is negligible and beam wandering can be ignored in the absence of an atmosphere. As a result, losses are attributed solely to diffraction. Assuming perfect alignment between the transmitting and receiving satellites, the transmittance can be modelled as \cite{Liu2022Composable}:

\begin{equation}
    T_{\mathrm{inter-satellite}} = 1 - \exp\left( -\frac{2r_a^2}{w^2(z)} \right),
\end{equation}
where $z$ is the link distance, $r_a$ is the radius of the receiver aperture on the receiver satellite. The beam radius, $w(z)$, as a function of the link distance is defined as:

\begin{equation}
    w(z) = w_0\left( 1 + \left[ \frac{\lambda z}{\pi w_0^2}   \right]^2    \right)^{\frac{1}{2}},
\end{equation}
where $w_0$ is the beam-waist radius and $\lambda$ is the wavelength.

\subsection{Satellite-to-Ground}
The main sources of loss in a satellite-to-ground link come from clouds, atmospheric turbulence, and atmospheric aerosols. These correspond to geometric losses from the link distance, scintillation losses from atmospheric turbulence, and scattering losses from atmospheric aerosols.

\begin{figure}[htp!]
    \centering
    \includegraphics[width=0.7\textwidth]{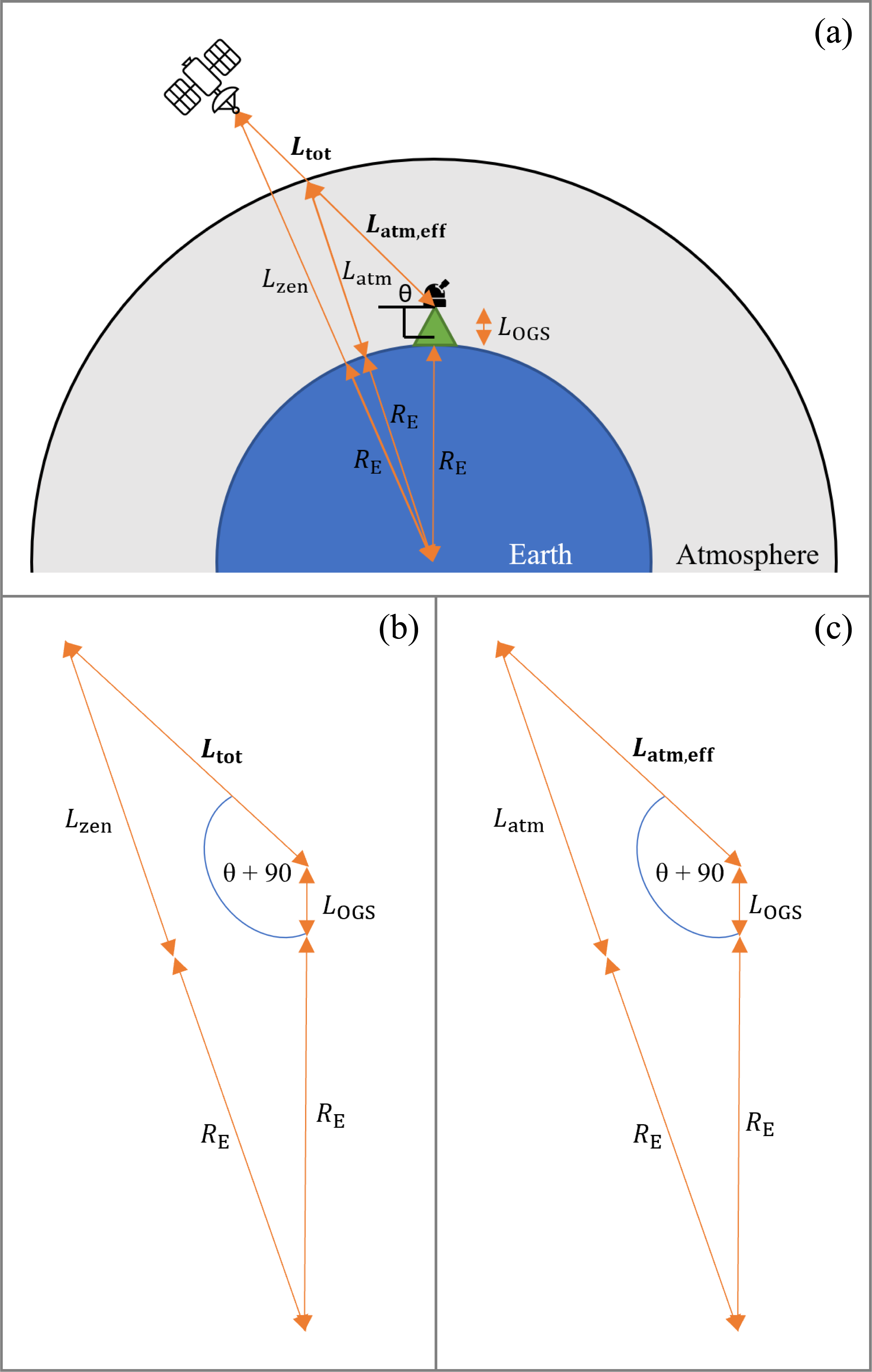}
    \caption[The satellite-to-ground channel model]{The satellite-to-ground channel model is shown in (a). Diagrams (b) and (c) shows the trigonometric determination of the total link distance ($L_{\mathrm{tot}}$) and effective atmosphere thickness ($L_{\mathrm{atm,eff}}$), respectively. $L_{\mathrm{atm}}$: atmosphere thickness at zenith (20~km), $L_{\mathrm{zen}}$: satellite altitude at zenith, $L_{\mathrm{OGS}}$: OGS altitude, $\theta$: elevation angle, $R_{\mathrm{E}}$: mean radius of the Earth (6371~km) \cite{sayat2024satellite}.}
    \label{fig:Satellite_to_Ground_Channel_Model}
\end{figure}

Atmospheric losses depend on the thickness of the atmosphere and the overall atmospheric mass through which the signal propagates. On average, 95\% of the total atmosphere mass is confined to the first 20 km from ground to zenith \cite{Liorni2019Satellite, Zuo2020Atmospheric}. The overall architecture of a satellite-to-ground link is shown in Figure~ \ref{fig:Satellite_to_Ground_Channel_Model}. Here, the satellite is assumed to be in LEO and passes over an OGS. The total link distance is calculated as \cite{sayat2024satellite}:

\begin{equation}
    \begin{aligned}
        L_{\mathrm{tot}} = & \left(R_{\mathrm{E}} + L_{\mathrm{zen}}\right)^2 + \left(R_{\mathrm{E}} + L_{\mathrm{OGS}}\right)^2 -2(R_{\mathrm{E}} + L_{\mathrm{zen}})(R_{\mathrm{E}} + L_{\mathrm{OGS}})\cos{(\alpha_1)}^\frac{1}{2} \\
        \alpha_1 = & \arcsin{\left[\cos{(\theta)}\frac{(R_{\mathrm{E}} + L_{\mathrm{OGS}})}{R_{\mathrm{E}} + L_{\mathrm{zen}}}\right]} + (90 - \theta),
    \end{aligned}
\end{equation}
where $R_\mathrm{E} = 6371$ km is the radius of the Earth, $L_{\mathrm{OGS}}$ is the altitude of the OGS, $L_{\mathrm{zen}}$ is the altitude of the satellite at zenith (90\degree$\,$ elevation angle), and $\theta > 0$ is the elevation angle. 

The effective atmosphere thickness the signal propagates through is calculated as \cite{sayat2024satellite}:

\begin{equation}
    \begin{aligned}
        L_{\mathrm{atm,eff}} = & (R_{\mathrm{E}} + L_{\mathrm{atm}})^2 + (R_{\mathrm{E}} + L_{\mathrm{OGS}})^2 -2(R_{\mathrm{E}} + L_{\mathrm{atm}})(R_{\mathrm{E}} + L_{\mathrm{OGS}})\cos{(\alpha_2)}^\frac{1}{2} \\
        \alpha_2 = & \arcsin{\left[\cos{(\theta)}\frac{(R_{\mathrm{E}} + L_{\mathrm{OGS}})}{R_{\mathrm{E}} + L_{\mathrm{atm}}}\right]} + (90 - \theta),
    \end{aligned}
\end{equation}
where $L_\mathrm{atm} = 20$ km is the atmosphere thickness from ground to zenith containing 95\% of the total atmospheric mass.

A satellite-to-ground link is characterised differently depending on the direction of transmission. It can either be a downlink (satellite-to-ground) or uplink (ground-to-satellite). This is because the link is not uniform and contains both the atmosphere and space environments. Uplinks are known to be more difficult as the signal is degraded by the atmosphere first, causing significant beam wandering and increase in scintillation by the time it reaches the satellite resulting in larger losses \cite{Aspelmeyer2003Long, Bourgoin2013Comprehensive, Bedington2017Progress,andrews2000scintillation, aly2020plane}.

Table~\ref{tab:vis_cn2} summarises `good' and `bad' atmospheric weather conditions used for the calculation of SKRs \cite{sayat2024satellite}.

\begin{table}[htp!]
    \centering
    \caption{Atmospheric Parameters in Satellite-to-Ground Links}
    \begin{tabular}{|l|r|r|}
        \hline
        \textbf{Parameter} & Good & Bad \\
        \hline
        \hline
        Atmospheric Visibility, $V$                     & 200~km  & 20~km \\
        Refractive Index Structure Parameter, $C_n^2$                     & $10^{-16} \; \mathrm{m}^{-2/3} $ & $10^{-12} \; \mathrm{m}^{-2/3}$ \\ 
        \hline
    \end{tabular}
    \label{tab:vis_cn2}
\end{table}

\newpage
\subsubsection{Satellite Downlink}
Geometric loss is defined as \cite{Aspelmeyer2003Long}:

\begin{equation}
    A_{\mathrm{geo}} = 10\log_{10}\left(\frac{L_{\mathrm{tot}}^2\lambda^2}{D_{\mathrm{t}}^2 D_{\mathrm{r}}^2} \frac{1}{T_{\mathrm{t}}(1 - L_{\mathrm{p}})T_{\mathrm{r}}}\right) [\mathrm{dB}],
    \label{losses_geo}
\end{equation}
where $\lambda$ is the wavelength, $D_\mathrm{t}$ is the transmitter aperture diameter, $D_\mathrm{r}$ is the receiver aperture diameter, $T_\mathrm{t}$ is the transmitter optics efficiency, $T_\mathrm{r}$ is the receiver optics efficiency, and $L_{\mathrm{p}}$ is the pointing loss that can be attributed to the inefficiency of an acquisition, pointing, tracking (APT) system of the OGS as well as beam wandering. Equation \ref{losses_geo} assumes that the receiver is in the far field ($L_\mathrm{tot} > \frac{D_{\mathrm{t}}D_{\mathrm{r}}}{\lambda}$), the transmitter is diffraction limited, and there is no attenuation from the atmosphere, enabling the addition of other atmospheric losses (scattering and scintillation losses).

Atmospheric aerosols give rise to scattering losses through Rayleigh scattering, Mie scattering, and geometrical scattering \cite{Kim2001Comparison}. At a wavelength of 1550 nm, the effects of Rayleigh scattering are negligible and the effects of geometrical scattering, from rain, snow, and hail, are ignored as satellite-to-ground CVQKD, in general, is not possible during these weather conditions as they could cause damage to the telescope in an OGS \cite{sayat2024satellite}. 

Mie scattering is defined through the Kruse and Kim model as \cite{Grabner2010Fog}:

\begin{equation}
    \label{scattering}
    \begin{aligned}
        A_{\mathrm{scat}} = 10\log_{\mathrm{10}}(e)(\frac{3.912}{V})(\frac{\lambda}{550})^{-p} \:\mathrm{[dB/km]}, \\
        p =  
        \begin{cases} 
            1.6 & V \geq 50 \: \mathrm{km} \\
            1.3 & 6 \: \mathrm{km} \leq V < 50 \: \mathrm{km} \\
            0.16V + 0.34 & 1 \:\mathrm{km} \leq V < 6 \:\mathrm{km} \\
            V - 0.5 & 0.5 \:\mathrm{km} \leq V < 1 \:\mathrm{km} \\
            0 & V < 0.5 \:\mathrm{km}
       \end{cases}.
    \end{aligned}
\end{equation}

Scintillation loss from atmospheric turbulence is defined as \cite{Giggenbach2008Fading}:

\begin{equation}
    \begin{aligned}
        A_{\mathrm{sci}} =  4.343\left(\mathrm{erf}^{-1} \left(2p_{\mathrm{thr}} - 1\right) \left[2\ln(\sigma_I^{2} + 1)\right]^\frac{1}{2} 
         -\frac{1}{2}\ln(\sigma_I^{2} + 1)\right) \quad[\mathrm{dB}],
    \end{aligned}
    \label{scintillation}
\end{equation}
where $\sigma_I$ is the scintillation index and $p_{\mathrm{thr}}$ is the probability that the received power is below the minimum required to register a signal (equivalent to the fraction of link outage time).

The scintillation index, assuming a spherical wave, is defined as:

\begin{equation}
    \begin{aligned}
        \sigma_{\mathrm{I}}^{2}(D_{\mathrm{r}})= \mathrm{exp}\bigg\{ \frac{0.20\sigma_R^{2}}{[1 + 0.18d^{2} + 0.20\left(\sigma_R^{2}\right)^{\frac{6}{5}}]^{\frac{7}{6}}} 
        + \frac{0.21\sigma_R^{2} [1 + 0.24(\sigma_R^{2})^{\frac{6}{5}}]^{-\frac{5}{6}}}{1 + 0.90d^{2} + 0.21d^{2}\left(\sigma_R^{2}\right)^{\frac{6}{5}}} \bigg\} - 1, 
    \end{aligned}
    \label{scintillation_index}
\end{equation}
where $\sigma_R$ is the Rytov variance and $d = D_{\mathrm{r}}\left( \frac{\pi}{2\lambda_{\mathrm{atm,eff}}}  \right)^2$. 

The Rytov variance is defined as: 

\begin{equation}
    \begin{aligned}
        \sigma_R^{2} = 2.25k^{\frac{7}{6}} \int_{0}^{L_{\mathrm{atm,eff}}} C_n^{2}(z)(L_{\mathrm{atm,eff}}-z)^{\frac{5}{6}}  \, dz
    \end{aligned}
    \label{rytov_variance},
\end{equation}
where $k = \frac{2\pi}{\lambda}$ is the wave number and $C_n^2$ is the refractive index structure parameter describing the intensity of turbulence in the atmosphere. 

$C_n^2$ can be empirically defined as \cite{Andrews2009Near}:

\begin{equation}
    \begin{aligned}
        C_n^2(h) &= 0.00594\left( \frac{w}{27} \right)^2 (h \times 10^{-5})^{10} \exp\left(-\frac{h}{1000}\right) \\
        &+ 2.7 \times 10^{-16}\exp\left( -\frac{h}{1500} \right) + A\exp\left( -\frac{h}{100}  \right),
    \end{aligned}
\end{equation}
for altitudes between 3 and 24 km (within the 20 km atmospheric thickness \cite{Liorni2019Satellite, Zuo2020Atmospheric}), where $w$ is the wind speed, $h$ is the altitude, and $A$ is the is the nominal value of $C_n^2$ at $h = 0$. Alternatively, $C_n^2$ can be defined qualitatively where $C_n^2 = 10^{-16}$ describes a low turbulent atmosphere, $C_n^2 = 10^{-14}$ a medium turbulent atmosphere, and $C_n^2 = 10^{-13}$ a high turbulent atmosphere \cite{Muhammad2005Channel}.

\subsubsection{Satellite Uplink}
In an uplink, losses are more prominent as the beam is exposed to the atmosphere first \cite{Aspelmeyer2003Long, Bourgoin2013Comprehensive, Bedington2017Progress}. This causes significant beam wandering and scintillation \cite{Dios2004Scintillation}. The uplink model can use most of the sources of loss found in a downlink with a few exceptions:

\begin{itemize}
    \item $L_{\mathrm{p, uplink}} > L_{\mathrm{p, downlink}}$ to account for greater beam wandering,
    \item $\sigma_{I, \mathrm{uplink}}^2 > \sigma_{I, \mathrm{downlink}}^2$ to account for larger scintillation loss.
\end{itemize}

The beam wandering arises from the difference in refractive index between the atmosphere and space. By propagating through the atmosphere first, the laser is deflected from its original path earlier and therefore will have greater deviation from the intended path to the satellite by the time it reaches the satellite altitude. This is represented as an increase in the pointing loss ($L_P$) in the satellite-to-ground model. However, this could be mitigated by acquisition, pointing, and tracking (APT) systems and optics which are more easily implemented in an OGS than a satellite \cite{Pirandola2021Satellite}. The same pointing loss, $L_P = 0.1$ is therefore used for both uplink and downlink SKR calculations. 

In contrast, for a given turbulence strength, quantified by the refractive index structure parameter ($C_n^2$), uplinks have greater scintillation indices ($\sigma_I$) compared to downlinks \cite{andrews2000scintillation, aly2020plane}. The difference between the uplink and downlink scintillation index increases as a function of the elevation angle. For a wavelength of 1550~nm and turbulence of $C_n^2 = 10^{-14} \; \mathrm{m}^{-2/3}$ (medium turbulence), the difference in scintillation index increases asymptotically to $\sigma_{I,\mathrm{uplink}} - \sigma_{I,\mathrm{downlink}} = 0.2$ \cite{aly2020plane}. For a weaker turbulence ($C_n^2 = 10^{-16} \; \mathrm{m}^{-2/3}$), it is expected that the difference between uplink and downlink scintillation indices would decrease. To prevent an underestimation of scintillation indices in uplinks, the following relation between uplink and downlink scintillation indices is used for all elevation angles:

\begin{equation}
    \sigma_{I, \mathrm{uplink}} = \sigma_{I, \mathrm{downlink}} + 0.2.
\end{equation}

\subsection{Sources of Noise}
In physical application, the excess noise will vary depending on the link. In an ideal system where coherent states are transmitted, received, and measured with a variance of 1 SNU, the excess noise represents unaccounted noise above this shot noise limit. For example, given the electronic noise is trusted, the excess noise in a transmitted shot noise limited coherent state with a measured variance of 1.2~SNU is 0.2~SNU. Table \ref{tab:excessnoise_in_links} shows different sources of excess noise that may be found in the link types modelled. Excess noise stems from coherent states that have undergone decoherence as well as measured photons which have not been taken into account.

\begin{table}[htp!]
    \centering
    \caption{Sources of Excess Noise}
    \begin{tabular}{|l|l|}
        \hline
        \textbf{Link Type} & Excess Noise Source \\
        \hline
        \hline
        Fibre                & Vibrations, heat \\
        Underwater           & Precipitates, small organisms, heat currents \\
        Inter-satellite      & Radiation  \\
        Satellite-to-ground  & Atmospheric aerosols, atmospheric turbulence \\
        \hline
    \end{tabular}
    \label{tab:excessnoise_in_links}
\end{table}

\section{Developing a Dynamic Global CVQKD Network}
A general network can be represented as a graph. Secret key distribution through a CVQKD network can be modelled as a graph problem where the secret key traverses through the graph to be shared between different nodes. The concept of the link capacity, which represents the maximum size of the secret key that can be transmitted in a link, is presented and can be used as a guide to determine optimum paths through a CVQKD network.

\subsection{Link Capacity}
The SKR is a valuable metric for showing the performance of a CVQKD protocol in a link. However, it does not include a key parameter; the time taken for transmitting coherent states and generating a secret key between nodes. The overall capacity of a link can be determined. The link capacity represents the maximum size of the secret key that can be transmitted in a link (note that the SKR is a lower bound and so the link capacity also becomes a lower bound). 

The link capacity can be presented as:

\begin{equation}
    C = \mathrm{SKR}\times t,
    \label{eqn:LinkCapacity}
\end{equation}
where $\mathrm{SKR}$ is the finite size limit SKR and $t$ is the period of time that the link can be used (positive SKR) to transmit the secret key. It is important to note that $t$ is not necessarily an arbitrary period of time. In satellite-to-ground links, it represents the usable period during a pass of a satellite over an OGS. In this case, there is a different SKR for each time increment as the satellite track varies in elevation angle and consequently varies in link distance. In contrast, $t$ has a single value in the case of fibre links as both nodes are stationary and there are no geometric dependencies other than a constant length.

In the case of secret key distribution through a CVQKD network, the link capacity acts as a guide to determine the optimum path, through nodes and links, through which a secret key can be distributed. For a practical multi-node distribution, the following link capacity conditions must be met:
\begin{itemize}
    \item The link capacity of subsequent links must be greater than the link capacity of the first link. This ensures that there are no links acting as a bottleneck during distribution, and
    \item All link capacities in a series of nodes and links must be greater than the size of the secret key being transmitted. This ensures that there is no information loss of the secret key during distribution. 
\end{itemize}

To show how the link capacity can be used, satellite-to-ground and inter-satellite links have been simulated.

\subsubsection{Satellite-to-Ground Link Simulation}
\begin{figure}[htp!]
    \centering
    \includegraphics[width=0.5\textwidth]{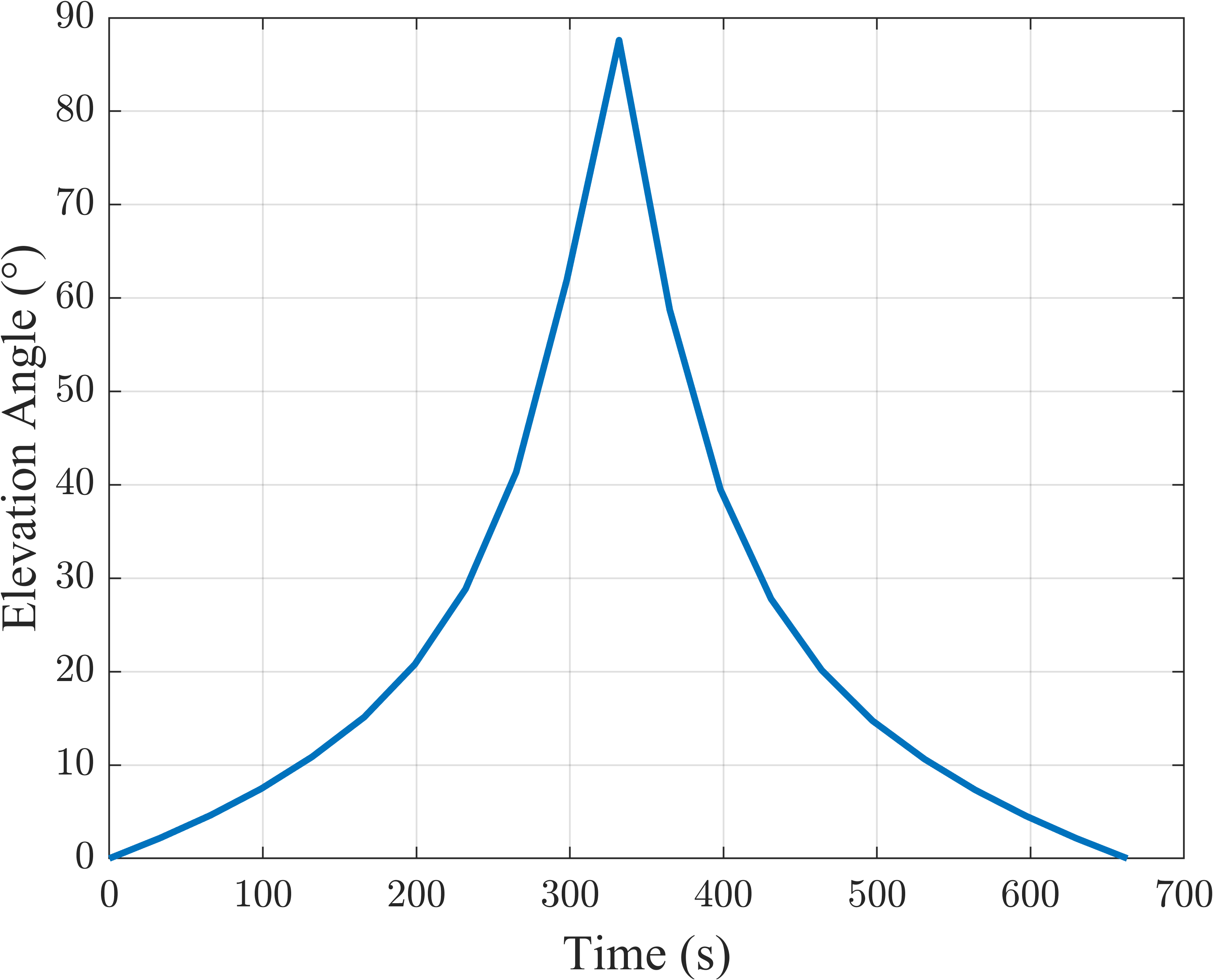}
    \caption[Elevation angles for an ISS pass over Mt. John Observatory]{Elevation angles for an ISS pass over Mt. John Observatory on 9th August 2022. The pass had a maximum elevation angle of 87.6\degree.}
    \label{fig:ISS_88_Pass}
\end{figure}

\paragraph{1. Simple Downlink}
Calculation of the link capacity has been performed between the University of Canterbury's Mt John Observatory (MJO) in New Zealand (Latitude = -43.9853\degree, Longitude = 170.4641\degree, Altitude = 1.029~km) and the International Space Station (ISS) for a downlink scenario \cite{sayat2024satellite}. Here, it was assumed that there is a CVQKD receiver at MJO and a CVQKD transmitter on the ISS. The ISS pass details were sourced from Gpredict \cite{Gpredict} for the 9th August 2022 and had a duration of 663~s with a maximum elevation angle of 87.6\degree (shown in Figure \ref{fig:ISS_88_Pass}).

\begin{figure}[htp!]
    \centering
    \includegraphics[width=0.6\textwidth]{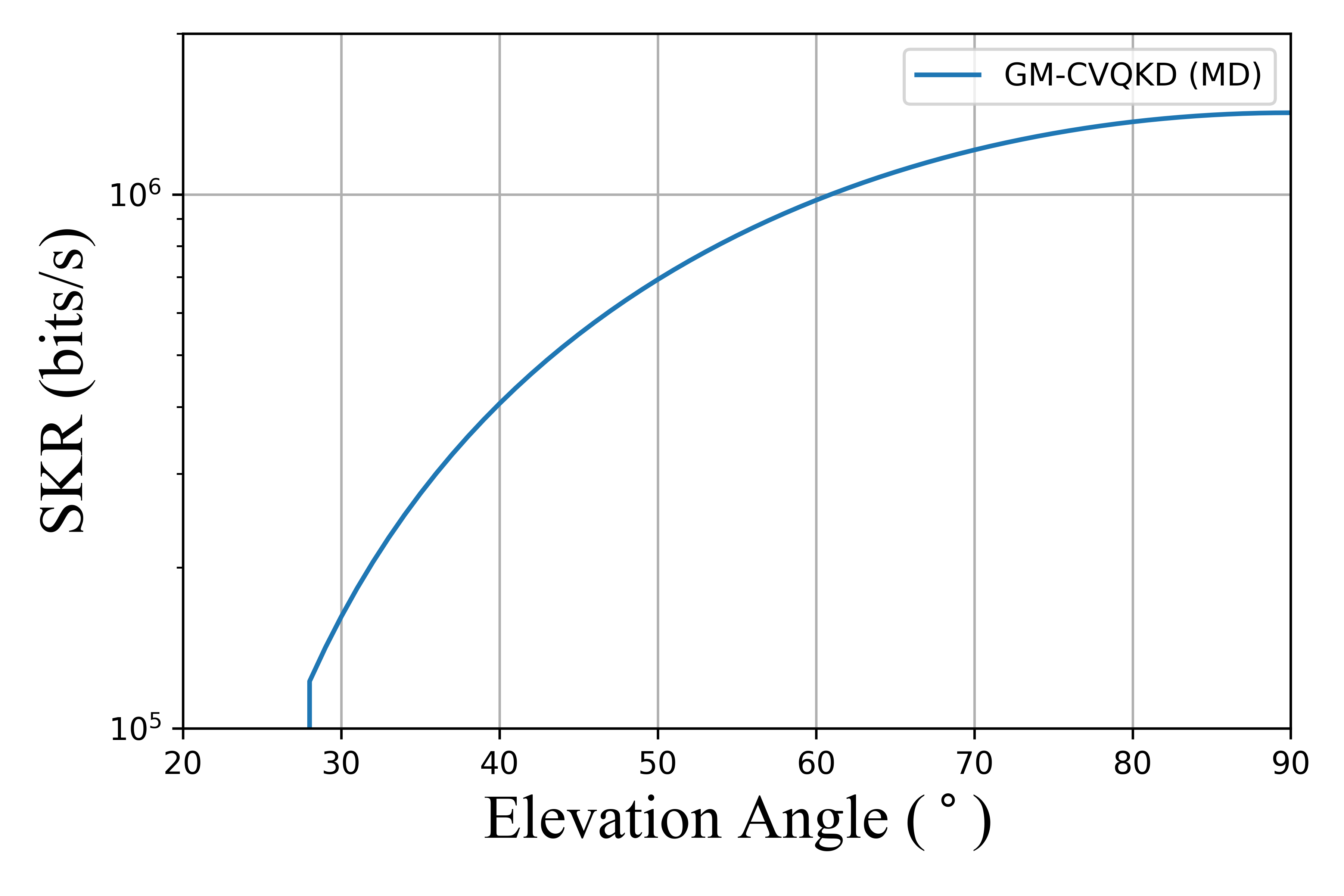}
    \caption[Downlink finite size limit SKRs for GM-CVQKD using MD reconciliation as a function of elevation angle.]{Downlink finite size limit SKRs for GM-CVQKD using MD reconciliation as a function of the elevation angle. $V_A = 5$ SNU, $\beta = 0.9$, $\xi$ = 0.03 SNU, $D_R = 1 \mathrm{m}$, $D_T = 0.3 \mathrm{m}$, $L_{\mathrm{zen}} = 408$ km, $L_{\mathrm{OGS}} = 1.029$ km.}
    \label{fig:STG_DownlinkSKR_Fin_Paper1}
\end{figure}

The calculation of the link capacity \cite{sayat2024satellite} is modified using a recent formulation of Gaussian modulated CVQKD (GM-CVQKD) \cite{Denys2021Explicit} for the calculation of the finite size limit SKR. In addition, an OGS altitude of $L_{OGS} =$ 1.029 km, good atmospheric conditions ($C_n^2 = 10^{-16} \mathrm{m}^{-2/3}$, $V = 200$ km), and ISS altitude of $L_{\mathrm{zen}} = 408$ km are used. Figure \ref{fig:STG_DownlinkSKR_Fin_Paper1} shows the finite size limit SKRs as a function of the elevation angle. 


Multiplying the calculated SKR by the time that the ISS is at a discrete elevation angle (Equation~\ref{eqn:LinkCapacity}) gives the maximum number of bits that can be transmitted during the downlink with GM-CVQKD with MD reconciliation. The elevation angle has been discretised with a resolution of $1\degree$ in Figure~\ref{fig:ISS_88_Pass} to determine the temporal position of the ISS. In this case, the link capacity is 155 Mbit if transmission were to occur during the pass when the elevation angle produces positive SKRs.

\paragraph{2. Uplink and Downlink through the Deep Space Network}
The importance of the link capacity is better demonstrated when there are multiple nodes and links used to distribute a secret key. For this case, a series of uplinks and downlinks are simulated using the locations of the Deep Space Network (DSN) with ground stations in Madrid (Latitude = 40.4298\degree, Longitude = -4.2489\degree), Goldstone (Latitude = 35.2828\degree, Longitude = -116.7840\degree), and Canberra (Latitude = -35.4025\degree, Longitude = 148.9829\degree) \cite{kegege2012three}, and the ISS (orbit altitude of $L_{\mathrm{zen}} = 408$ km). 

\begin{figure}[htp!]
    \centering
    \includegraphics[width=0.55\textwidth]{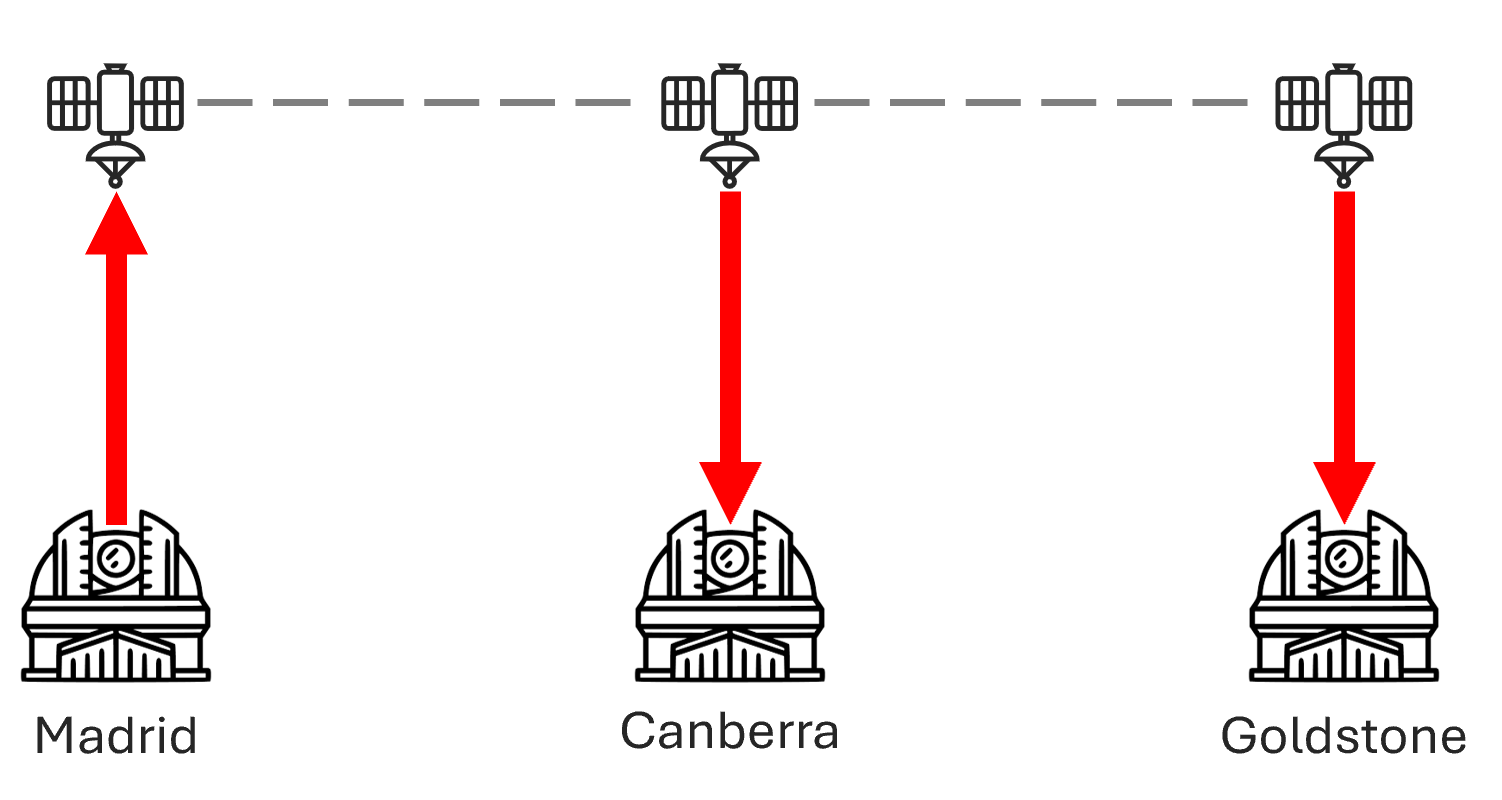}
    \caption{Schematic of the uplink and downlinks scenario for the ISS passing over the three DSN ground station sites.}
    \label{fig:DSN_ISS_Flyby}
\end{figure}

\begin{figure}[htp!]
    \centering
    \includegraphics[width=0.5\textwidth]{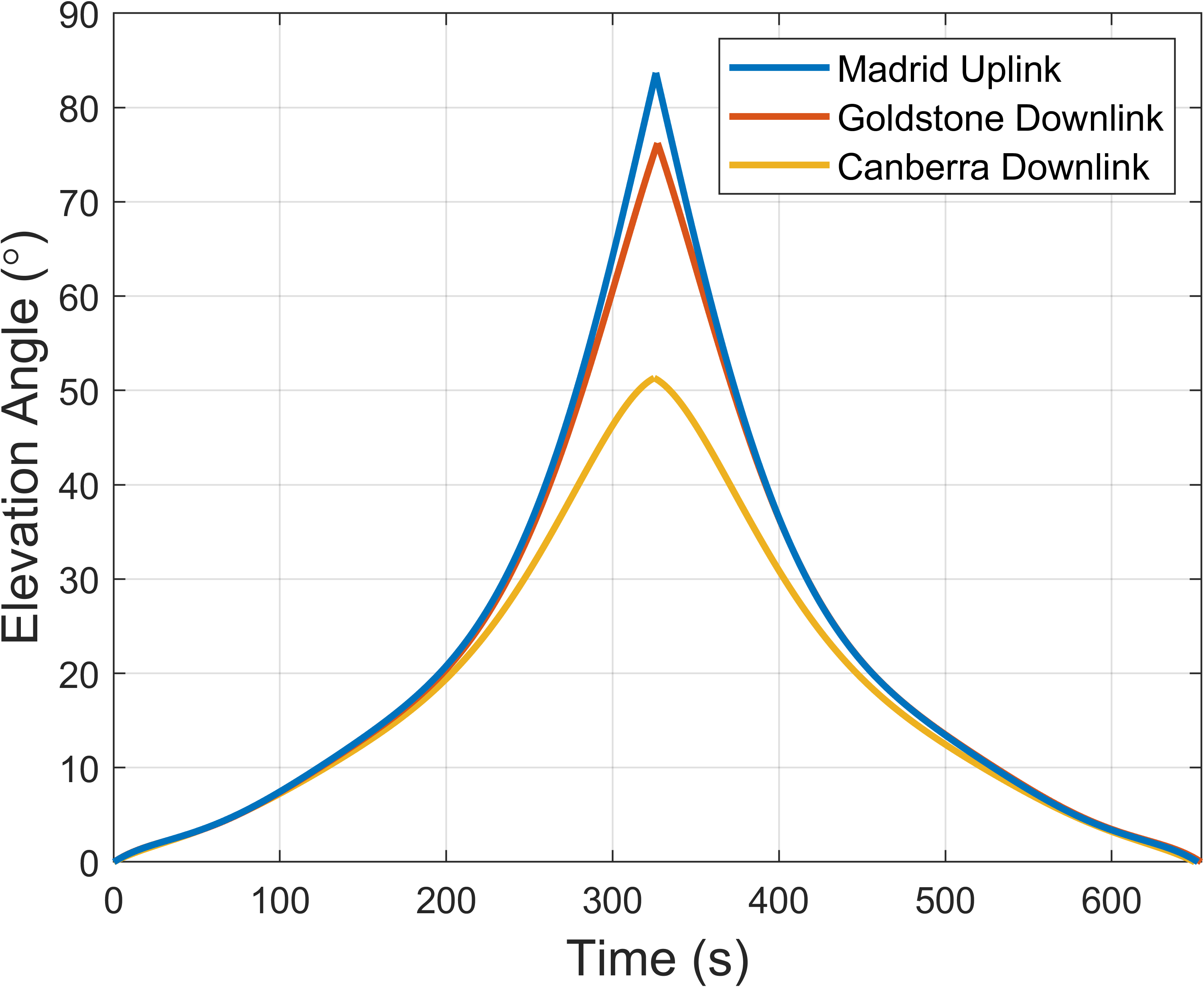}
    \caption{Elevation angles of ISS passes over Madrid, Goldstone, and Canberra DSN ground stations.}
    \label{fig:DSN_Passes}
\end{figure}

A scenario was modelled using CVQKD where the distribution of the secret key with an uplink from Madrid to the ISS, downlink from the ISS to Canberra, and downlink from the ISS to Goldstone (Figure~\ref{fig:DSN_ISS_Flyby}), where each transmission occurs when the ISS passes over the field of view of each ground station. The passes of the ISS over the three ground stations were again estimated with Gpredict \cite{Gpredict}, and are presented in Figure \ref{fig:DSN_Passes}.

\begin{figure}[htp!]
    \centering
    \includegraphics[width=0.9\textwidth]{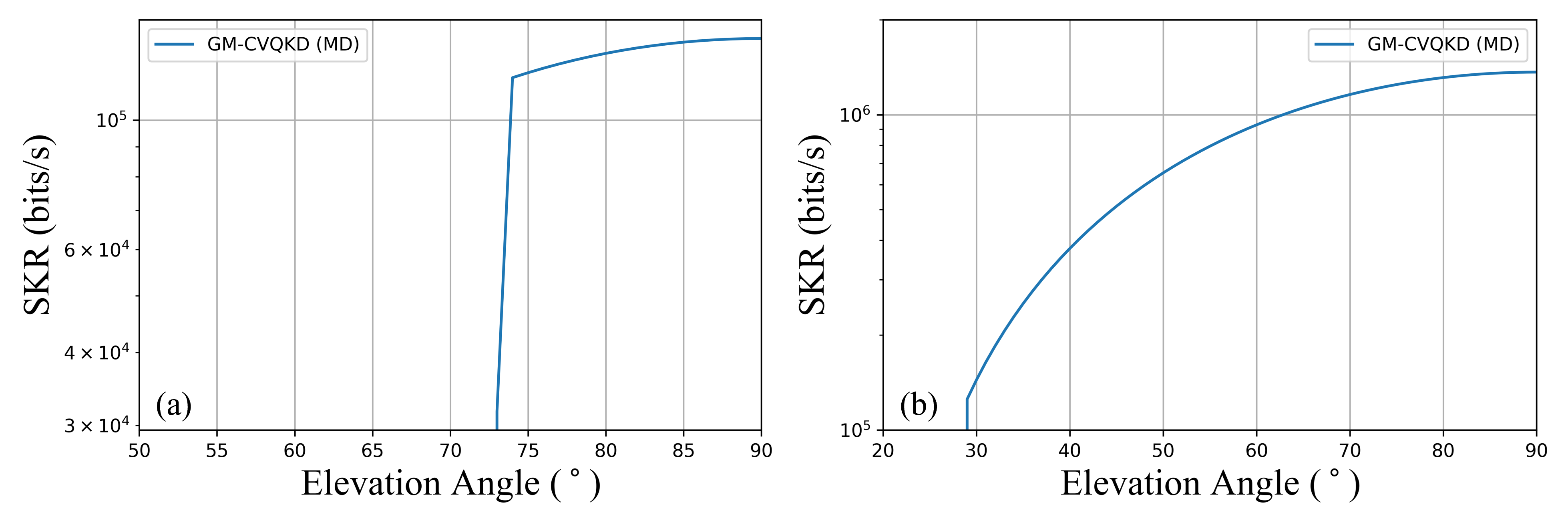}
    \caption{(a) Uplink and (b) downlink finite size limit SKRs for GM-CVQKD using MD reconciliation as a function of elevation angle. Using MLC-MSD reconciliation did not produce positive SKRs. $D_R = 0.3 \; \mathrm{m}$, $D_T = 1 \; \mathrm{m}$, $L_{\mathrm{zen}} = 408$ km.}
    \label{fig:STG_UplinkDownlinkSKR_Fin}
\end{figure}

The corresponding uplink and downlink finite size limit SKRs used to calculate the link capacities are the SKR profiles shown in Figure \ref{fig:STG_UplinkDownlinkSKR_Fin}. Similar to the simple downlink simulation, the SKR at each elevation angle was multiplied by the duration that the ISS spends at that elevation angle, where the elevation angle has been discretised into $1\degree$ increments. 

Table \ref{tab:UpDownLinkCapacities} shows the maximum uplink and downlink link capacities. It can be seen that although the Madrid pass reaches the highest elevation angle of all the passes (Figure \ref{fig:DSN_Passes}), it does not yield the largest link capacity. This is because the uplink SKR profile (Figure \ref{fig:STG_UplinkDownlinkSKR_Fin}a) only produces positive SKRs above an elevation angle of approximately $73\degree$, and the ISS pass spends less time at larger elevation angles. Similarly, the Canberra pass does not yield the smallest link capacity because the downlink SKR profile (Figure \ref{fig:STG_UplinkDownlinkSKR_Fin}b) has a smaller minimum elevation angle (approximately $29\degree$), and the ISS pass spends more time at smaller elevation angles.

\begin{table}[htp!]
\caption{Link Capacity for each Uplink and Downlink} 
\label{tab:UpDownLinkCapacities}
\begin{center}       
\begin{tabular}{|l|r|r|r|} 
\hline
\rule[-1ex]{0pt}{3.5ex}  & Madrid Uplink & Canberra Downlink & Goldstone Downlink  \\
\hline
\rule[-1ex]{0pt}{3.5ex}  Link Capacity (MBit) & 3.71 & 61.53 & 121.25  \\
\hline 
\end{tabular}
\end{center}
\end{table}

Given that the secret key size (SK) transmitted is less than or equal to 3.71 MBit, the scenario is feasible according to the link capacity criteria as

\begin{enumerate}
    \item Subsequent links have a greater link capacity than the first link:\\ $C_{\mathrm{Canberra, downlink}}, C_{\mathrm{Goldstone, downlink}} > C_{\mathrm{Madrid, uplink}}$, and
    \item All link capacities are greater than the size of the secret key being transmitted:\\
    $C_{\mathrm{Madrid, uplink}}, C_{\mathrm{Canberra, downlink}}, C_{\mathrm{Goldstone, downlink}} > \mathrm{SK}$.
\end{enumerate}

\subsubsection{Inter-Satellite Link Simulation}
\label{sec:inter-satellitelinkSimulations}
A satellite pass over an OGS with a large field of view (and therefore a large range of available elevation angles) to support large SKRs and link capacities is infrequent in most locations. A promising solution is the use of inter-satellite links. In this case, a satellite with the secret key from an OGS uplink can pass the secret key via a chain of satellites until a satellite within an acceptable field of view over the receiving ground station is reached.

\begin{figure}[htp!]
   \centering
   \includegraphics[width=0.7\textwidth]{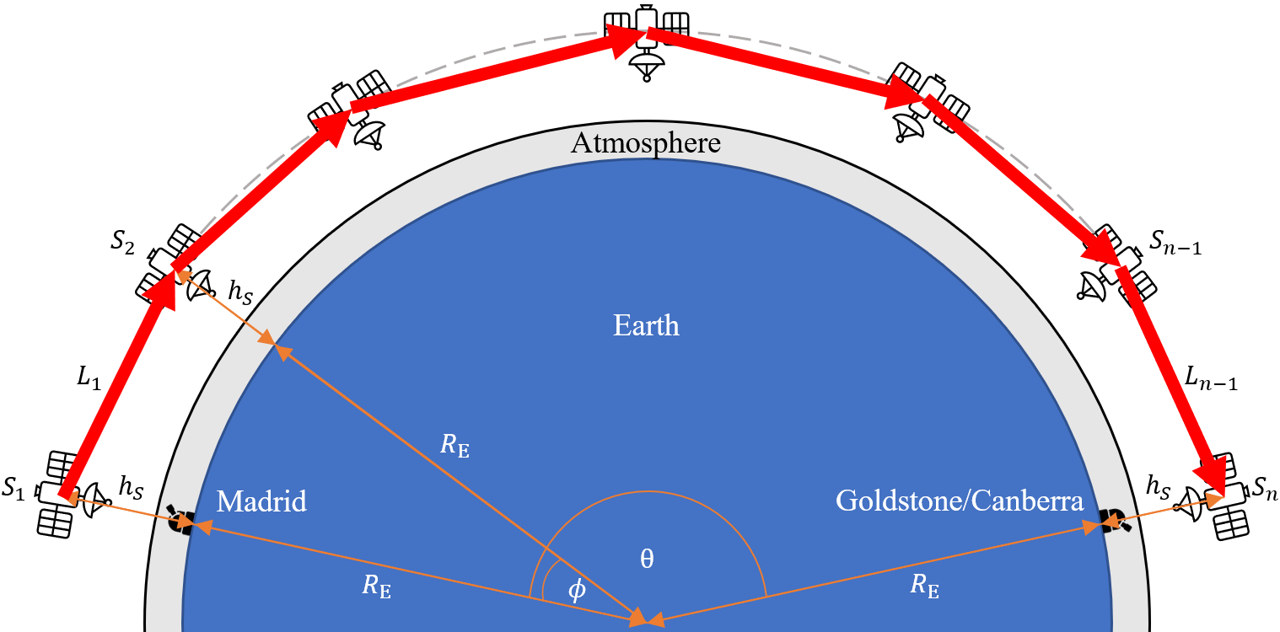}
   \caption[A schematic of secret key distribution between Madrid and Goldstone/Canberra using inter-satellite links.]{A schematic of secret key distribution between Madrid and Goldstone/Canberra using inter-satellite links. $S:$ Satellite, $L$: Inter-satellite link, $n$:~Number of satellites in the chain of inter-satellite links, $R_\mathrm{E}$: Earth radius, $h_S$: Satellite orbit altitude, $\theta$: Central angle between the two ground stations (Madrid and Goldstone/Canberra), $\phi$: Central angle between two satellites with an inter-satellite link.}
   \label{fig:inter_satellite}
\end{figure}

The minimum number of inter-satellite links between Madrid and Goldstone, and Madrid and Canberra can be found by assuming the link capacity values in Table \ref{tab:UpDownLinkCapacities}. This scenario is presented in Figure \ref{fig:inter_satellite} with the calculated distances and central angles shown in Table \ref{tab:IS_Measurements}. The chain of satellites were assumed to be at the same altitude as the ISS i.e. $h_s = 408$ km. 

\begin{table}[htp!]
\centering
\caption{Distances and Central Angles between Madrid and Goldstone/Canberra}
\begin{tabular}{|l|r|r|r|}
\hline
\multicolumn{1}{|c|}{Link} & \multicolumn{1}{c|}{\begin{tabular}[c]{@{}c@{}}Ground \\ Distance (km)\end{tabular}} & \multicolumn{1}{c|}{\begin{tabular}[c]{@{}c@{}}Distance\\  @ 408 km \\ Altitude (km)\end{tabular}} & \multicolumn{1}{c|}{\begin{tabular}[c]{@{}c@{}}Central \\ Angle (\textdegree)\end{tabular}} \\ \hline
Madrid to Goldstone        & 9,200                                                                                & 9,789                                                                                              & 83                                                                               \\ \hline
Madrid to Canberra         & 17,600                                                                               & 18,727                                                                                             & 158                                                                              \\ \hline
\end{tabular}
\label{tab:IS_Measurements}
\end{table}

\begin{figure}[htp!]
    \centering
    \includegraphics[width=0.6\textwidth]{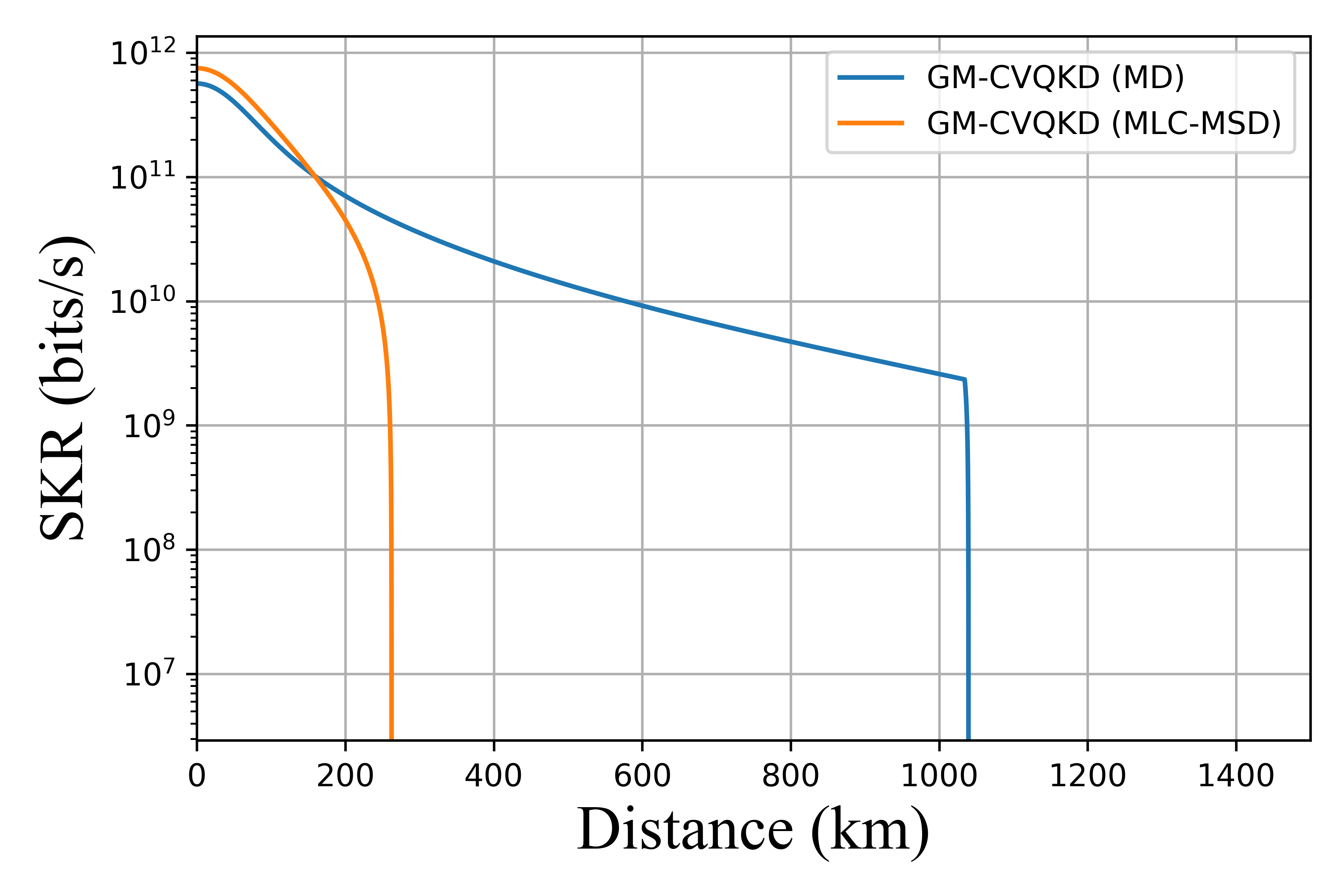}
    \caption[Inter-satellite based finite size limit SKRs for GM-CVQKD using MLC-MSD and MD reconciliation.]{Inter-satellite based finite size limit SKRs for GM-CVQKD using MLC-MSD and MD reconciliation. $r_a~=~w_0~=~0.2 \; \mathrm{m}$.}
    \label{fig:ISSKRs_f}
\end{figure}

The inter-satellite link finite size limit SKR profile in Figure \ref{fig:ISSKRs_f} is used to calculate the link capacity. The SKR at 1000~km is approximately 3000~Mbit/s for GM-CVQKD with MD reconciliation. Each link in the inter-satellite chain only needs to be greater than the Madrid uplink link capacity which is 3.71~MBit. This requires a transmission time greater than 1.2~ms and results in a central angle between adjacent satellites of $\phi = 8.46\degree$. This would correspond to 9-10 links and 10-11 satellites in the Madrid to Goldstone inter-satellite chain, and 18-19 links and 19-20 satellites in the Madrid to Canberra inter-satellite chain.

An inter-satellite link distance of 1000~km between adjacent satellites at an altitude of 408~km has the entire laser link at an altitude above the atmosphere (which has a thickness of 20~km \cite{Liorni2019Satellite, Zuo2020Atmospheric}). For longer distances, made possible by larger apertures, it is important that the laser does not propagate through the atmosphere to avoid atmospheric losses. This is ensured by optimising values for $\phi$ and $L_n$ so that the laser is above 20 km.

\subsection{Global CVQKD Network: A Dynamic Graph Problem}
The different types of links that could be used for establishing a global CVQKD network are classified here and shown in Table \ref{tab:LinkClassifications}. The geometry of the link is dependent on whether a node is stationary (fixed location) or moving (varying location). Stationary-to-stationary node links are the simplest but are confined to relatively small distances due to the relatively high attenuation in fibre links. Stationary-to-moving node links require an APT system that enables the capture and tracking of the moving node. Moving-to-moving node links requires APT systems with high resolution and fast tracking as there is relative motion between the nodes.

As the link capacity is time dependent, the three types (stationary--stationary, stationary--moving, moving--moving) are classified here into static or dynamic links (Table \ref{tab:LinkClassifications}). A static link is where the link capacity is only dependent on the transmission time. This occurs when the link is spatially fixed and contains no moving transmitters and receivers. Dynamic links occur when the link capacity is dependent on both the transmission time and the link geometry between the transmitter and receiver as the link distance between them can vary.

Uniform links are defined here as when the channel between the transmitter and receiver can be thought of as isotropic. For example, inter-satellite links ideally have the vacuum of space between the transmitter and receiver \cite{Liu2022Composable}, and the channel in inter-submarine (underwater) links can be modelled as water with an average chlorophyll concentration \cite{Xu2016Underwater,Kong2017Security}. In contrast, non-uniform links have a non-isotropic channel where the link capacity is direction-dependent. This is appropriate for satellite-to-ground links where the SKR and link capacity differ between uplinks and downlinks because the channel contains both the atmosphere and vacuum of space (Figure \ref{fig:GlobalQuantumCVQKDNetwork}). In addition, the channel changes during a satellite pass since the total link distance and effective atmosphere thickness changes.

It should be noted that if a constellation of satellites were developed where the relative positions between each satellite were fixed through active propulsion, then inter-satellite links are static (as modelled in Section \ref{sec:inter-satellitelinkSimulations}). This is because the relative distance between each satellite remains fixed. In addition, submarine-to-ground links would require an OGS to be near the coast with direct access to the sea to form a uniform link with a submarine where the only environment is sea-water.

\begin{table}[!htp]
\caption{Classification of Different Links} 
\label{tab:LinkClassifications}
\begin{center}
\scalebox{1}{
\begin{tabular}{|l|l|l|l|} 
\hline
\rule[-1ex]{0pt}{3.5ex} \textbf{Link} & \textbf{Geometry} & \textbf{Static/Dynamic} & \textbf{Uniform/Non-Uniform}  \\
\hline
\rule[-1ex]{0pt}{3.5ex}  Fibre & Stationary-Stationary & Static  &  Uniform \\
\hline
\rule[-1ex]{0pt}{3.5ex}  Satellite-to-Ground & Stationary-Moving  & Dynamic &  Non-Uniform \\
\hline
\rule[-1ex]{0pt}{3.5ex}  Submarine-to-Ground & Stationary-Moving & Dynamic &  Uniform \\
\hline
\rule[-1ex]{0pt}{3.5ex}  Satellite-to-Submarine & Moving-Moving & Dynamic  &  Non-Uniform \\
\hline
\rule[-1ex]{0pt}{3.5ex}  Inter-Satellite & Moving-Moving & Dynamic/Static &  Uniform \\
\hline
\rule[-1ex]{0pt}{3.5ex}  Inter-Submarine & Moving-Moving & Dynamic & Uniform  \\
\hline
\end{tabular}}
\end{center}
\end{table}

\begin{figure}[htp!]
    \centering
    \includegraphics[width=0.6\textwidth]{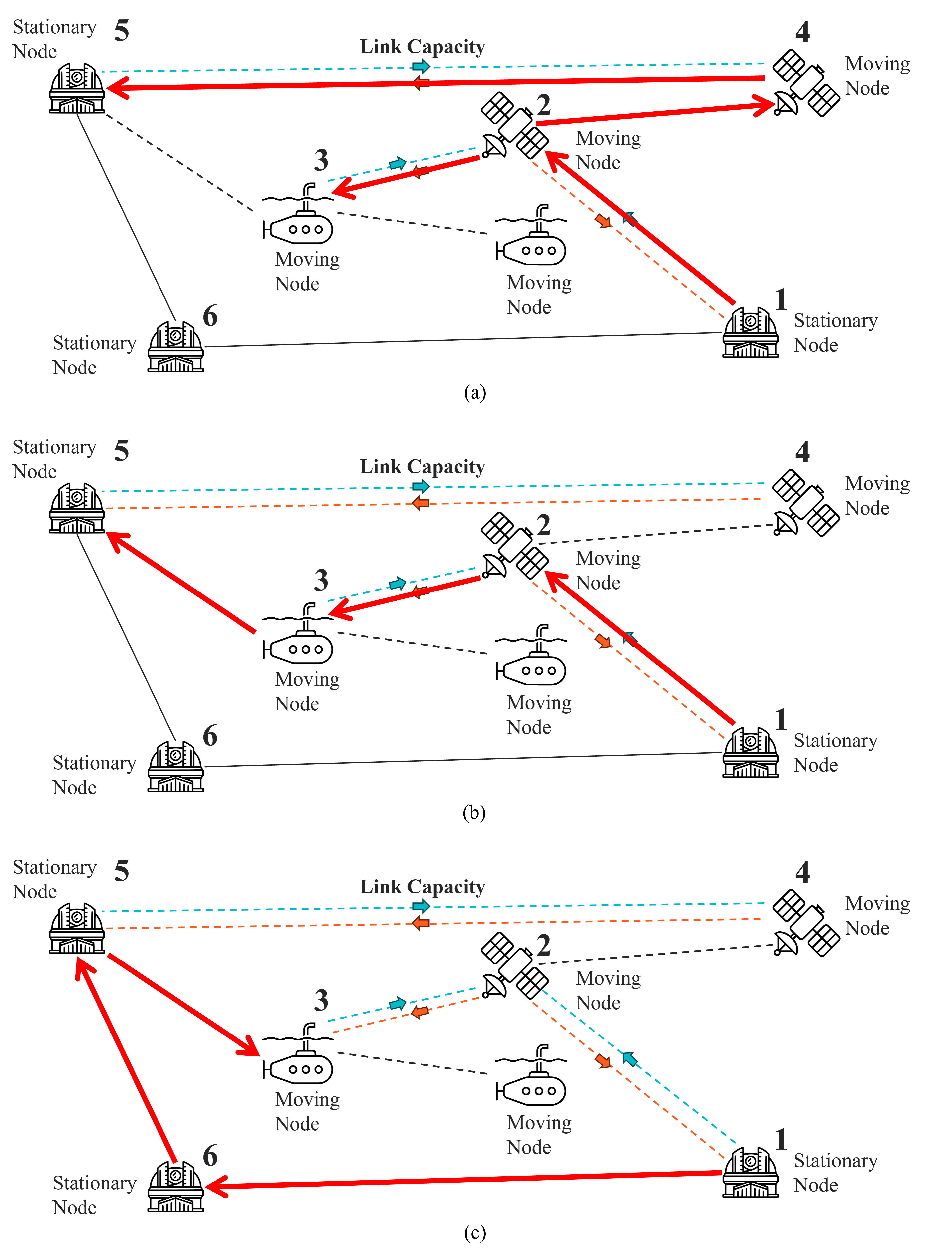}
    \caption{Possible pathways to distribute a secret key between an OGS (1), submarine (3), and another OGS (5).} 
    \label{fig:CVQKD_Graph_Traversing}
\end{figure}

The different nodes and links can be represented graphically as shown in Figure \ref{fig:CVQKD_Graph_Traversing}, where the nodes comprise the different CVQKD transmitters and receivers, and the edges comprise the different link types. The weights represent the link capacity. Figure \ref{fig:CVQKD_Graph_Traversing} also shows a hypothetical scenario where a secret key is to be distributed between an OGS (1), a submarine (3), and another OGS (5). In this case, there can be three pathways as shown in Figure \ref{fig:CVQKD_Graph_Traversing}.

The availability and use of different links depends on factors such as the weather, and spatiotemporal availability of nodes, \cite{Bennet2020Australia}. For a graph, the link capacity can be used as the ``cost" to traverse through different links throughout the network. Shortest path algorithms could be employed, such as Dijkstra's algorithm \cite{gupta2016applying}, to find the optimum path to distribute the secret key to the desired nodes. Here, the optimum path would be the links with the largest link capacities while satisfying the link capacity conditions for multi-node transmissions.

In a general QKD network architecture (Figure \ref{fig:NetworkArchitecture}), the network manager would track the position of nodes and monitor the dynamic link capacities in real-time. In addition, the network manager would determine and monitor the optimum path between nodes for CVQKD based on the link capacities at any given time should a request arise for secret key distribution.

\section{Discussion}
The SKR of different CVQKD in different link types have been analysed in the asymptotic and finite size limit. In the finite size limit, GM-CVQKD with MD reconciliation outperforms all other protocols. So far, CVQKD has been implemented in fibre and short-distance horizontal free-space. However, a full-stack implementation in all link types with post-processing is yet to be realised in addition to physical implementation in other link types such as satellite-to-ground, inter-satellite, and underwater links.


The simulated inter-satellite scenario is idealised (Section \ref{sec:inter-satellitelinkSimulations}). It assumes that adjacent satellites are within line-of-sight of each other with adequate APT systems for alignment. In addition, it assumes instantaneous/negligible processing time of the secret key. In reality, this is not the case and processing will cause a delay in transmitting a secret key between satellites. Sources of delay may include the time taken to align satellite transmitter and receiver apertures, decode the received secret key and re-transmit to another satellite (in the case of a trusted satellite), relay the secret key through quantum repeaters. The satellites in the chain are also assumed to be trusted. This gives assumed security within each satellite node where Eve cannot access it. However, if any of the satellite are not trusted, quantum repeaters \cite{furrer2018repeaters,dias2020quantum} would be required to maintain security. The scenario assumes well-positioned and consistently spaced satellites within the chain. In reality, the satellites orbit with different trajectories and a smaller number of satellites may be required when transmitting the secret key as the satellites move. The satellite trajectories are predictable and the route could be optimised.

As developments in CVQKD progress, a global quantum network secured by CVQKD can be modelled virtually. Real-time locations of nodes and possible links between nodes can be modelled and implemented in such a virtual network. The spatiotemporal feasibility of CVQKD in different link types would depend on the link capacity, which is dependent on the dynamic SKR for links involving moving nodes, such as in the simulations performed for satellite-to-ground and inter-satellite links. Real-time weather and environmental conditions could also be included to simulate potential adversities that could be present in the different links. In addition, a governing network architecture could be implemented, simulating a CVQKD secured global quantum network. The network manager would monitor parameters related to CVQKD and the weather/environment across the network. In particular, the link capacity to determine the optimum path when nodes and users request a secret key. The virtual network would act as a foundation for physically deploying CVQKD in a global quantum network when the necessary developments are made such as experimentally validated CVQKD in different link types, integration of different links, and controllability of links and nodes.

In developing a global quantum network secured with general QKD, it is important to use the already existing resources such as current national networks around the world. Since these networks are predominantly DVQKD-based, CVQKD would be an addition to the network; creating a hybrid global quantum network which uses both CVQKD and DVQKD. The network would still require a governing network architecture. Therefore, this architecture would require compatibility with both CVQKD, DVQKD, and all other technologies found in the different links and nodes in existing QKD networks. In addition, future implementations e.g. for CVQKD, would also need to be compatible.

The application of shortest path algorithms for finding the optimum path for secret key distribution requires frequent updates \cite{tanizawa2016routing}. Since there are moving nodes and dynamic links, in addition to varying environmental conditions, the optimum path would vary with time. This would depend on a range of variables such as the real-time positions of nodes, link distances, and the losses involved in the links which are environment-dependent. The positions of nodes and therefore links must be tracked in combination with environmental parameters such as the weather (e.g. turbulence) and chlorophyll concentrations to determine real-time link capacities across the network. This could be achieved by keeping track of satellite TLEs (two line elements), submarine positions and the chlorophyll concentration of the waters they are in when revealed, the locations of OGSs in use, and therefore the dynamic link capacities across the network.

The distribution of secret keys through different paths across a network (Figure \ref{fig:CVQKD_Graph_Traversing}) assumes a trusted node framework (trusted relay implementation) in which the secret key can simply be received and transmitted through different nodes as they are assumed to be resilient to eavesdroppers. However, there is the case where certain nodes cannot be trusted, and therefore passing a secret key through them poses a security risk. The underlying assumption in QKD is that any unaccounted for noise/loss detected above an upper bound threshold can be attributed to the presence of an eavesdropper \cite{djordjevic2019quantum,liu2021homodyne,lo2008quantum}. In the case of remote nodes such as satellites, increases in noise can be detected but its source cannot always be determined. For example, a spike in noise can be attributed to system malfunction (e.g. overheating of detectors) or an actual eavesdropper. To combat this problem, quantum repeaters that are compatible with CVQKD must be used \cite{furrer2018repeaters,dias2020quantum}. In this case, a secret key passing through a node is not measured but simply relayed onto the next node. Since the secret key is not measured at a node, then any technical problems leading to noise variations in a fully functioning node with a quantum repeater can be blamed solely on an eavesdropper tampering with the node. Although quantum repeaters are suitable for this endeavour, they are currently in development \cite{ruihong2019research} and have yet to be used for satellite applications. 

A method for secret key distribution is to build up storage of secret keys over time at certain locations (nodes) where users can then request the use of secret keys \cite{Cao2022Evolution}. This can be achieved with satellite-to-ground CVQKD where secret keys are generated and stored during good weather conditions. A network of different nodes and links could improve upon this by helping store secret keys by providing alternative paths when certain links are not available (from bad channel parameters e.g. bad weather conditions) through spatial diversity.  The use of shortest path algorithms in combination with link capacities could provide optimum paths for secret key distribution in this endeavour.


The links established using OGS, satellite, and submarine nodes were used in the context of a CVQKD network. The addition of high-altitude platforms such as (space) planes, UAVs, and drones adds another dimension for faster real-time CVQKD in localised areas \cite{conrad2021drone, sidhu2021advances}. The controllability of these platforms could allow for more targeted CVQKD avoiding reliance on the trajectories of nodes such as satellites, providing more spatial diversity and interconnectedness. In addition, they have the potential to operate in all weather-conditions and can connect ground and satellite platforms together \cite{sidhu2021advances}. Moreover, the development of transportable OGSs (TOGSs) \cite{Fuchs2013DLR} provides further spatial diversity on the ground. This could give rise to more dynamic links and easier access to secret keys.

\section{Conclusions}
A study of the structure of different QKD networks around the world showed different numbers of architectural layers in each network. The different layers were fit into a general three-layer QKD network architecture which consists of an application layer, control and management layer, and infrastructure layer representing how the different networks could be combined into an overarching global QKD network. The existing QKD network layers would have to be adapted to fit within the three-layer QKD network architecture. The use of an SDN controller would help ease the integration of new nodes as well as integration into existing telecommunication networks.

In general, a larger and more positive SKR for longer link distances can be achieved in inter-satellite links followed by satellite-to-ground, fibre, and underwater links in order. This is based on the analysis of different transmittance models inherent to each link in combination with constant values of modulation variance, reconciliation efficiency, and excess noise. Since the performance of CVQKD differs in each link type, spatiotemporal diversity of nodes is beneficial to combat situations where dynamic links become unfavourable to distribute a secret key. 


Different links which can be used in a global QKD network were classified and such a network was presented as a graph problem. The link capacity which considers a dynamic SKR based on a dynamic link (consisting of moving nodes) and its dynamic parameters was developed to calculate the maximum lower bound of information that can be stored in the secret key in a link for CVQKD. This can be used as a routing metric to determine the feasibility of secret key distribution between different nodes across a network. In addition, the link capacity, in combination with search algorithms, can help determine the optimum path through a CVQKD network that hosts a plethora of dynamic and non-dynamic links. 

\medskip
\textbf{Supporting Information} \par 
Supporting Information is available from the Wiley Online Library or from the author.

\medskip
\textbf{Acknowledgements} \par 
M. T. Sayat is a University of Auckland Doctoral Scholar. This research is supported by A*STAR under Project No. C230917009, and Q.InC Strategic Research and Translational Thrust.

\medskip
\textbf{Conflict of Interest} \par
The authors declare no conflict of interest.

\newpage
\section*{Appendix A: SKR as a Function of Distance}

\begin{figure}[htp!]
    \centering
    \includegraphics[width=\textwidth]{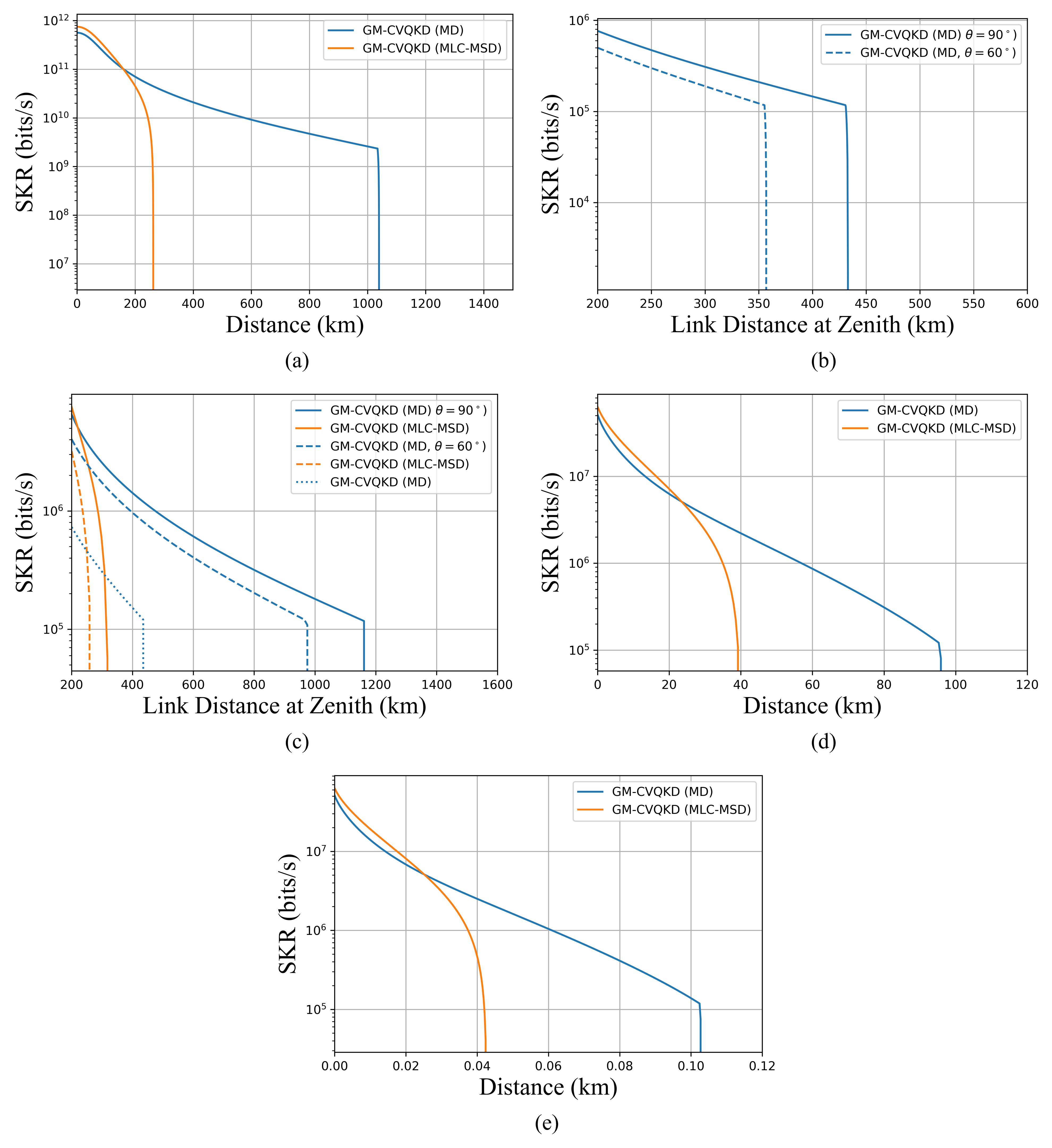}
    \caption{Finite size limit SKRs for GM-CVQKD using MD and MLC-MSD reconciliation in the different links studied: (a) Inter-satellite, $r_a = w_0 = 0.2 \; \mathrm{m}$; (b) Uplink, $D_R = 0.3 \; \mathrm{m}$, $D_T =~1 \; \mathrm{m}$. Using MLC-MSD reconciliation did not produce positive SKRs; (c) Downlink, $D_R = 1 \; \mathrm{m}$, $D_T = 0.3 \; \mathrm{m}$; (d) Fibre-based; (e) Underwater with pure sea water.} 
    \label{fig:AppAGroup}
\end{figure}

Analysis of the different parameters that affect different links show that there are multiple ways to increase the SKR. The parameters inherent to each link are based on the SKR calculation in Section \ref{sec:SecretKeyRate} and the transmittance/loss equations found in Section \ref{sec:QKDNetworks}.

In general, the following results in more positive and larger SKRs:
\begin{itemize}
    \item Larger transmittance by mitigating losses,
    \item Smaller excess noise from better characterisation of sources of noise,
    \item Optimised modulation variance from more accurate characterisation of links and the covariance matrix,
    \item Higher correlation coefficient from more accurate estimation of modulation variance,
    \item Higher reconciliation efficiency from high performance information reconciliation codes,
    \item Higher repetition rate from laser systems, and
    \item Smaller $\delta n_{\mathrm{privacy}}$ from more accurate estimates of the Holevo bound and more symbols sent.
\end{itemize}

\paragraph{Fibre links} The main parameter in fibre links is the attenuation factor. Decreasing this leads to larger transmittances and therefore more positive and higher SKRs.

\paragraph{Satellite-to-Ground/Submarine Links} Favourable atmospheric conditions and optimisation of a CVQKD system lead to more positive and larger SKRs. In this case:
\begin{itemize}
    \item Increasing the transmitter and receiver aperture diameter as well as the transmitter and receiver optics efficiencies increases the transmittance and therefore the overall achievable link distance.
    \item Decreasing the pointing loss increases the maximum link distance and therefore satellite orbit altitude. This requires more accurate APT systems.
    \item Increasing the OGS elevation increases the maximum link distance and therefore satellite orbit altitude for which a positive SKR is possible. This results from the decrease in attenuation from a decrease in link distance and effective atmosphere thickness as Bob is effectively closer to Alice.
    \item Operating in atmospheric conditions where there is greater visibility, smaller refractive index structure parameter ($C_n^2$), and smaller scintillation increases the overall link distance. This results from the decrease in attenuation due to less atmospheric scattering, turbulence, and beam wandering.
\end{itemize}

\paragraph{Inter-satellite links} The main parameters influencing inter-satellite links are the receiver aperture and the beam waist radius. Increasing these values leads to larger transmittances and therefore more positive and higher SKRs.

\paragraph{Underwater links} The absorption and scattering coefficients are the main parameters which affect the transmittance in underwater links. These coefficients are influenced by the chlorophyll concentration in the water. Having a smaller concentration of chlorophyll leads to smaller absorption and scattering coefficients and consequently a larger transmittance which leads to more positive and larger SKRs.

\paragraph{}Overall, the calculation and parameter analysis of SKRs show that inter-satellite links allow the longest link distances followed by satellite-to-ground links, fibre links, and underwater links in descending order. Although inter-satellite links (Figure \ref{fig:AppAGroup}a) show smaller positive SKRs for slightly smaller distances than downlinks (Figure \ref{fig:AppAGroup}c), this is because smaller apertures were used in the calculation to represent typical aperture sizes in space. For the same sized apertures, inter-satellite links achieve larger positive SKRs for larger distances compared to satellite-to-ground links. It can also be seen that MD reconciliation outperforms MLC-MSD reconciliation by producing larger positive SKRs at larger link distances.

\medskip

%

\bibliographystyle{MSP}
\bibliography{References}

\end{document}